\documentclass[10pt,twocolumn,english,prd,superscriptaddress,nofootinbib,preprintnumbers,showpacs,floatfix]{revtex4-2}
\usepackage{graphicx,color}
\usepackage{amsmath}
\usepackage{amssymb}
\usepackage{hyperref}
\usepackage[utf8]{inputenc}
\usepackage[english]{babel}
\usepackage{epsfig}
\usepackage{subfigure}
\usepackage{wasysym}
\usepackage{color,xcolor}
\usepackage{amsmath}
\usepackage{bm}
\usepackage{epsfig}
\usepackage{amsfonts}
\usepackage{dcolumn}
\usepackage{float}

\usepackage{makecell} 
\usepackage{comment}

\usepackage{cases}

\hypersetup{colorlinks=true, linkcolor=blue, citecolor=green}

\usepackage{dcolumn}% Align table columns on decimal point
\usepackage{bm}% bold math
\usepackage{ifpdf}
\usepackage{hyperref}
\usepackage{float}
\usepackage{bm}
\usepackage{xcolor,color,graphicx,graphics}
\usepackage[OT1]{fontenc}
\usepackage{latexsym,amssymb,amsmath,amsfonts}
\usepackage{makeidx}
\usepackage{epsfig}
\usepackage{epstopdf}
\usepackage{mathrsfs}
\hypersetup{colorlinks=true, linkcolor=blue, citecolor=green}
\usepackage{enumerate}
\usepackage{xcolor}
 \usepackage{multirow}
\usepackage{tikz}
\newcommand{\orcidicon}{%
	\begin{tikzpicture}
	\draw[lime, fill=lime] (0,0) 
		circle [radius=0.16] 
		node[white] {{\fontfamily{qag}\selectfont \tiny ID}};
	\draw[white, fill=white] (-0.0625,0.095) 
		circle [radius=0.007];
	\end{tikzpicture}	\hspace{-2mm}
}
\newcommand\orcidEdnaldo{{\href{https://orcid.org/0000-0002-9388-8373}{\orcidicon}}}
\newcommand\orcidFrancisco{{\href{https://orcid.org/0000-0002-9388-8373}{\orcidicon}}}
\newcommand\orcidManuel{{\href{https://orcid.org/0000-0001-8586-0285}{\orcidicon}}}
\newcommand\orcidTarciso{{\href{https://orcid.org/0009-0007-0450-2672}{\orcidicon}}}
\newcommand\orcidHenrique{{\href{https://orcid.org/0000-0001-7565-4277}{\orcidicon}}}
\newcommand\orcidLuis{{\href{https://orcid.org/0009-0009-4322-6484}{\orcidicon}}}
\newcommand\orcidJorde{{\href{https://orcid.org/0009-0001-3344-2986}{\orcidicon}}}
\newcommand\orcidDaniela{{\href{https://orcid.org/0009-0002-3463-0417}{\orcidicon}}}
\begin{document}
%========================================================

\title{Black bounce solutions via nonminimal scalar–electrodynamic couplings} 

%=================================================================
\author{Daniela S. J. Cordeiro\orcidDaniela\!\!}         \email{fc52853@alunos.ciencias.ulisboa.pt}        
\affiliation{Instituto de Astrof\'{i}sica e Ci\^{e}ncias do Espa\c{c}o, Faculdade de Ci\^{e}ncias da Universidade de Lisboa, Edifício C8, Campo Grande, P-1749-016 Lisbon, Portugal}

	\author{Ednaldo L. B. Junior\orcidEdnaldo\!\!} \email{ednaldobarrosjr@gmail.com}
\affiliation{Faculdade de F\'{i}sica, Universidade Federal do Pará, Campus Universitário de Tucuruí, CEP: 68464-000, Tucuruí, Pará, Brazil}
\affiliation{Programa de P\'{o}s-Gradua\c{c}\~{a}o em F\'{i}sica, Universidade Federal do Sul e Sudeste do Par\'{a}, 68500-000, Marab\'{a}, Par\'{a}, Brazil}

%========================================================
	\author{Jos\'{e} Tarciso S. S. Junior\orcidTarciso\!\!}
 \email{tarcisojunior17@gmail.com}
\affiliation{Faculdade de F\'{\i}sica, Programa de P\'{o}s-Gradua\c{c}\~{a}o em 
F\'isica, Universidade Federal do 
 Par\'{a},  66075-110, Bel\'{e}m, Par\'{a}, Brazil}

%=================================================================
	\author{Francisco S. N. Lobo\orcidFrancisco\!\!} \email{fslobo@ciencias.ulisboa.pt}
\affiliation{Instituto de Astrof\'{i}sica e Ci\^{e}ncias do Espa\c{c}o, Faculdade de Ci\^{e}ncias da Universidade de Lisboa, Edifício C8, Campo Grande, P-1749-016 Lisbon, Portugal}
\affiliation{Departamento de F\'{i}sica, Faculdade de Ci\^{e}ncias da Universidade de Lisboa, Edif\'{i}cio C8, Campo Grande, P-1749-016 Lisbon, Portugal}

%========================================================
	\author{Jorde A. A. Ramos\orcidJorde\!\!}
 \email{jordealves@ufpa.br}
\affiliation{Faculdade de F\'{\i}sica, Programa de P\'{o}s-Gradua\c{c}\~{a}o em 
F\'isica, Universidade Federal do 
 Par\'{a},  66075-110, Bel\'{e}m, Par\'{a}, Brazil}
 
%=================================================================
	\author{Manuel E. Rodrigues\orcidManuel\!\!}
	\email{esialg@gmail.com}
	\affiliation{Faculdade de F\'{\i}sica, Programa de P\'{o}s-Gradua\c{c}\~{a}o em 
F\'isica, Universidade Federal do 
 Par\'{a},  66075-110, Bel\'{e}m, Par\'{a}, Brazil}
\affiliation{Faculdade de Ci\^{e}ncias Exatas e Tecnologia, 
Universidade Federal do Par\'{a}\\
Campus Universit\'{a}rio de Abaetetuba, 68440-000, Abaetetuba, Par\'{a}, 
Brazil}

%=================================================================

\author{Luís F. Dias da Silva\orcidLuis\!\!} 
        \email{fc53497@alunos.fc.ul.pt}
\affiliation{Instituto de Astrof\'{i}sica e Ci\^{e}ncias do Espa\c{c}o, Faculdade de Ci\^{e}ncias da Universidade de Lisboa, Edifício C8, Campo Grande, P-1749-016 Lisbon, Portugal}

%=================================================================

 \author{Henrique A. Vieira\orcidHenrique\!\!} \email{henriquefisica2017@gmail.com}
\affiliation{Faculdade de F\'{i}sica, Programa de P\'{o}s-Gradua\c{c}\~{a}o em F\'{i}sica, Universidade Federal do Par\'{a}, 66075-110, Bel\'{e}m, Par\'{a}, Brazill}

%-----------------------------------------------------------------
\date{\LaTeX-ed \today}
%%%%%%%%%%%%%%
%========================================================
%%%%%%%%%%%%%%%%%%%%%%%%%%%%%%%%%%%%%%%%%%%%%%%%%%%%%%%%%%%%%%%%

%%%%%%%%%%%%%%%%%%%%%%%%%%%%%%%%%%%%%%%%%%%%%%%%%%%%%%%%%%%%%%%%
%========================================================
\begin{abstract}
%========================================================
Black-bounce (BB) solutions generalize the spacetimes of black holes, regular black holes, and wormholes, depending on the values of certain characteristic parameters. In this work, we investigate such solutions within the framework of General Relativity (GR), assuming spherical symmetry and static geometry.
It is well established in the literature that, in order to sustain such geometries, the source of Einstein’s equations in the BB context can be composed of a scalar field $\varphi$ and a nonlinear electrodynamics (NLED). In our model, in addition to the Lagrangian associated with the scalar field in the action, we also include an interaction term of the form $W(\varphi)\mathcal{L}(F)$, which introduces a nonminimal coupling between the scalar field and the electromagnetic sector. Notably, the usual minimal coupling configuration is recovered by setting $W(\varphi) = 1$.
In contrast to approaches where the function $W(\varphi)$ is assumed a priori, here we determine its functional form by modeling the radial dependence of the derivative of the electromagnetic Lagrangian as a power law, namely $\mathcal{L}_F(r) \sim F^n$. This approach enables us to determine $W(r)$ directly from the obtained solutions.
We apply this procedure to two specific geometries: the Simpson–Visser-type BB solution and the Bardeen-type BB solution, both analyzed in the purely magnetic ($q_m \neq 0$, $q_e = 0$) and purely electric ($q_m = 0$, $q_e \neq 0$) cases. In all scenarios, we find that these BB spacetime solutions can be described with a linear electrodynamics, which is a noteworthy result.
Furthermore, we examine the regularity of the spacetime through the Kretschmann scalar and briefly discuss the associated energy conditions for the solutions obtained.

\end{abstract}

%========================================================
%=================================================================
\maketitle
%=================================================================
\def\HMS{{\scriptscriptstyle{\rm HMS}}}
%========================================================
%\bigskip
%\hrule
\tableofcontents
%\bigskip
%\hrule
%========================================================
%\parindent0pt
%\parskip7pt
%========================================================

%%%%%%%%%%%%%%%%%%%%%%%%%%%%%%%%%%%%%%%%%%%%%%%%%%%%%%%%%%%%%%%%
\section{Introduction}\label{sec1}
%%%%%%%%%%%%%%%%%%%%%%%%%%%%%%%%%%%%%%%%%%%%%%%%%%%%%%%%%%%%%%%%

The theory of General Relativity (GR), formulated by Einstein in 1916~\cite{Einstein:1916vd}, remains the most successful description of gravitation, showing outstanding consistency with both experimental tests and astrophysical observations. The first exact solution of Einstein’s field equations — and arguably its most iconic — was obtained shortly thereafter by Karl Schwarzschild~\cite{Schwarzschild}. This solution describes a compact object surrounded by a boundary of no return, the event horizon. In Schwarzschild’s formulation, the mass was the sole defining parameter, leading to a vacuum solution valid outside the object. Today, however, it is well established that black hole (BH) solutions in GR may be fully characterized by up to three parameters: mass, electric charge, and angular momentum, in line with the ``no-hair'' theorem \cite{Israel:1967PR}.
In recent years, unprecedented progress in high-precision astrophysical measurements has renewed interest in GR, largely because of the experimental confirmation of some of its most remarkable predictions. Noteworthy milestones include the first direct detection of gravitational waves in 2015 by the LIGO and Virgo Collaborations~\cite{LIGOScientific:2016aoc, LIGOScientific:2017ync}, generated by black hole mergers, as well as the imaging of the event horizons of the supermassive black holes at the centers of M87 and the Milky Way, accomplished by the Event Horizon Telescope~\cite{EventHorizonTelescope:2019dse, EventHorizonTelescope:2022wkp}.

Despite its remarkable success, GR faces several fundamental challenges. A prominent example is the unavoidable central singularity in the Schwarzschild solution, located at $r = 0$, where spacetime curvature becomes infinite and the classical laws of GR break down, resulting in geodesic incompleteness. This type of singular behavior is not unique to the Schwarzschild solution but also appears in other classical BH solutions, regardless of charge or rotation, making it a general feature of black hole spacetimes in GR.
To address this limitation and develop more complete models, various strategies have been explored, either by modifying the theory of gravity itself or by coupling Einstein’s equations to additional fields. A straightforward approach involves incorporating electromagnetism as a matter source, enabling the study of charged solutions and their distinctive characteristics. This combination generates a rich spectrum of new solutions and provides a more comprehensive framework for exploring the interaction between gravitational and electromagnetic fields under extreme astrophysical and cosmological conditions.

In this context, so-called regular black holes (RBHs) emerge as solutions that are free of singularities~\cite{Ansoldi:2008jw}. The first such model was introduced by Bardeen in 1968~\cite{Bardeen}, featuring an event horizon but no divergences in spacetime. Decades later, Ayon-Beato and García (1999) demonstrated that Bardeen's solution can be derived as an exact solution of GR coupled to nonlinear electrodynamics (NLED), with the regularization parameter interpreted as the magnetic charge of a monopole~\cite{Ayon-Beato:2000mjt}, later extended to include the electric case~\cite{Rodrigues:2018bdc}. In general, some RBH solutions possess a de Sitter or Minkowski-type core at their center~\cite{Culetu:2015cna,Simpson:2019mud,Bronnikov:2024izh}.
NLED was originally motivated by the desire to eliminate the singularities associated with point charges and the divergence of their intrinsic energy in classical Maxwell theory. The foundational work that led to NLED, describing the behavior of electromagnetic fields at high intensities, was proposed by Born and Infeld in 1934~\cite{Born:1933pep,Born:1934gh}. Building on this framework, several classes of NLED solutions were subsequently developed, and their implications explored in a variety of contexts~\cite{Heisenberg:1936nmg,Plebanski,Kruglov:2014hpa,Kruglov:2014iwa,Kruglov:2014iqa,Bandos:2020jsw,Harko:2013xma,Harko:2013wka,Harko:2013aya}.

Building on this foundation, several RBH models have been developed by minimally coupling nonlinear electrodynamics (NLED) to the equations of motion of GR. This coupling has expanded the class of RBH solutions and offered new perspectives for investigating BH physics. Within this framework, several notable models with significant theoretical implications stand out~\cite{Bronnikov:2000vy,Dymnikova:2004zc,Balart:2014cga,Culetu:2014lca}. The presence of NLED in the spacetime geometry directly influences photon propagation~\cite{Novello:1999pg,Habibina:2020msd,Toshmatov:2021fgm,dePaula:2024yzy} and modifies the shape and size of BH shadows~\cite{Stuchlik:2019uvf,Allahyari:2019jqz,Kruglov:2020tes}. Moreover, the thermodynamic properties of these solutions have been extensively investigated using various approaches~\cite{Breton:2004qa,Myung:2007xd,Ma:2015gpa,Kruglov:2016ymq,Fan:2016hvf}.
RBH solutions have also been explored in the context of modified theories of gravity~\cite{Junior:2015fya,Rodrigues:2015ayd,Rodrigues:2016fym,deSousaSilva:2018kkt,Rodrigues:2019xrc,Tangphati:2023xnw,Junior:2023ixh,Junior:2024xmm}. These studies highlight the central role of NLED in the search for singularity-free solutions and highlight its importance in understanding the interplay between spacetime and electromagnetic fields in extreme astrophysical and cosmological regimes. For a more comprehensive discussion of the topic, we refer the reader to the recent review in~\cite{Lan:2023cvz}.

In addition to the RBH solutions, a new type of singularity-free configuration, known as the \textit{black bounce} (BB), was recently proposed by Simpson and Visser in 2018~\cite{Simpson:2018tsi}\footnote{Although this solution gained widespread attention with this work, an earlier analysis of the corresponding metric was already presented in \cite{Visser:1997yn}.}. The BB solution avoids the central singularity through a smooth modification of the radial coordinate, effectively introducing a bounce that keeps the radius of the spherical area nonzero. Specifically, the original proposal inserts a regularization parameter $a$ into the metric, replacing $r \to \sqrt{a^{2} + r^{2}}$. This generalization creates a versatile framework for gravitational analysis, as varying the parameter $a$ can yield different scenarios: black holes, regular black holes, or traversable wormholes (uni- or bidirectional).  
The latter correspond to solutions predicted by GR, representing theoretical structures that can connect different regions of the same universe or even distinct universes. Like RBHs, these spacetimes are free of singularities, featuring instead a central throat. Wormholes were initially proposed by Einstein and Rosen as non-traversable solutions~\cite{PhysRev.48.73}, and their properties have since been extensively studied, including the possibility of traversable geometries~\cite{Bronnikov:1973fh,Ellis:1973yv,Morris:1988cz,Barcelo:1999hq,Barcelo:2000zf,Visser:2003yf,Lobo:2005us,Lobo:2007zb,Cardoso:2016rao,Bronnikov:2017sgg,Lobo:2017cay,Blazquez-Salcedo:2020czn,Churilova:2021tgn,Konoplya:2021hsm}.

Research on BB-type solutions has addressed various aspects, including the regularity of spacetime, analysis of causal structures and the energy conditions \cite{Lobo:2020ffi}, wormhole geometries \cite{Lobo:2020kxn,Berry:2020tky}, gravitational lensing effects~\cite{Nascimento:2020ime,Tsukamoto:2020bjm,Cheng:2021hoc,Tsukamoto:2021caq,Zhang:2022nnj}, and possible observational signatures \cite{Guerrero:2021ues,Jafarzade:2021umv, Jafarzade:2020ova,Jafarzade:2020ilt,Yang:2021cvh,Bambhaniya:2021ugr,Ou:2021efv,Guo:2021wid,Wu:2022eiv,Tsukamoto:2022vkt}. Generalizations of the BB model to rotating configurations were also investigated \cite{Mazza:2021rgq,Xu:2021lff}, extending its applicability to more realistic astrophysical scenarios. In addition, solutions based on string clouds were found~\cite{Rodrigues:2022rfj,Yang:2022ryf}.
The BB solutions are not limited to the use of NLED as a source of matter. In several cases, exact solutions in GR were obtained by minimal couplings between NLED and scalar fields, where the regularization parameter was interpreted as the magnetic charge \cite{Bronnikov:2022bud,Canate:2022gpy,Rodrigues2023,Pereira:2023lck}. More recently, BB solutions supported by purely electric charges have been proposed~\cite{Lima:2023arg,Alencar:2024yvh} and by dyonic configurations, combining magnetic and electric charges~\cite{Junior:2025sjr}. In addition to the spherically symmetric spacetime, versions with cylindrical symmetry were also analyzed~\cite{Bronnikov:2023aya,Lima:2022pvc}. In parallel, BB solutions in modified theories of gravity were studied in detail~\cite{Junior:2022zxo,Junior:2023qaq,Junior:2024vrv,Junior:2024cbb,Rois:2024qzm,Rois:2025czu}.
Finally, we emphasize the general formalism presented in~\cite{Alencar:2025jvl}, which was developed for the systematic construction of BB solutions in static and spherically symmetric spacetimes.

Based on the above, there is a growing interest in both RBH and BB solutions, mainly due to their non-singular nature, which offers a promising route to addressing the central singularity problem that plagues classical BH spacetimes. In general, the construction of these solutions relies on the inclusion of NLED as a matter source, which provides the necessary modifications to the energy-momentum content to smooth out singularities. While this approach frequently yields regular geometries, it is not without important limitations. A prominent issue is the absence of a proper Maxwell limit in the weak-field region for many of these solutions, including Bardeen's RBH itself. This discrepancy is at odds with astrophysical observations, where electromagnetic fields predominantly display linear behavior in asymptotically flat regions, and any manifestations of NLED appear to be restricted to the immediate vicinity of compact objects~\cite{Mignani:2016fwz,Ejlli:2020yhk}.

Another important aspect, as briefly mentioned above, is that photon propagation in NLED can be affected by the so-called effective metric~\cite{Novello:1999pg}, which arises from the combination of the background spacetime metric with additional terms depending on the NLED Lagrangian density. However, these effective metrics do not always preserve fundamental properties, such as regularity, nor do they necessarily reduce to the classical Schwarzschild or Reissner–Nordström solutions when the parameters are appropriately chosen. This limitation introduces challenges in modeling and interpreting astrophysical observations, including the detailed analysis of black hole shadows. Furthermore, the thermodynamic behavior of NLED solutions remains an open problem: to date, there is no fully unified formulation of the first law of black hole thermodynamics for regular geometries, such as Bardeen's solution~\cite{Ma:2014qma,Zhang:2016ilt}.

In addition to satisfying the energy conditions, theories of NLED must meet further requirements to ensure physical consistency. In particular, Shabad and Usov~\cite{Shabad:2011hf} proposed two fundamental principles: (i) causality, requiring that elementary excitations on a background field do not propagate faster than the speed of light in vacuum, and (ii) unitarity, requiring that the propagator residue be non-negative to avoid probability violations. However, magnetic solutions with a regular center, including both BHs and solitons, inevitably violate these criteria near the center.  
Another challenge concerns the electrodynamic Lagrangian supporting the solutions. For regular electrically charged RBH or BB configurations, the function ${\cal L}(F)$ may lack a simple analytic form and can even be multivalued, i.e., taking more than one value for the same $F$~\cite{Bronnikov:2000vy,Rodrigues:2018bdc,Alencar:2024yvh}. This raises questions about the consistency of the matter-field description. Moreover, NLED-supported solutions may suffer from dynamical stability issues, depending on the type of perturbation considered, as discussed in several studies~\cite{Moreno:2002gg,Breton:2014nba,Toshmatov:2019gxg,Nomura:2020tpc,DeFelice:2024seu,DeFelice:2024ops}.  
Given these limitations, it is plausible that linear electrodynamics (LED) scenarios could mitigate some of these issues, offering a conceptually simpler and physically more tractable framework for probing gravity in extreme regimes.

In this work, we follow a procedure analogous to that presented in \cite{RBHLE}, where the authors construct solutions for regular BHs in static and spherically symmetric spacetimes, coupling the characteristic matter fields of these geometries within the framework of GR. Here, in addition to a Lagrangian associated with the scalar field, we introduce an interaction term of the form $W(\varphi){\cal L}(F)$, which implements a nonminimal coupling between the scalar field and the electromagnetic field. This extended structure enables the analytic reconstruction of the matter functions that sustain the geometry, while also ensuring the global regularity of the solution. Notably, in the particular case where $W(\varphi) = 1$, the standard configuration with minimal coupling between the scalar and electromagnetic fields is recovered, highlighting the generality and flexibility of the framework.

Although the study in \cite{RBHLE} focused on the regular spacetime of the Bardeen solution, taken as a representative example, the authors demonstrated that it is possible to obtain a matter source described by LED within the context of RBHs. Importantly, the method they developed does not depend on the specific properties of the Bardeen model, making it applicable to a broader class of regular geometries.
In the present work, we generalize this approach to the BB framework and examine the conditions on $W(\varphi)$ and ${\cal L}(F)$ that permit the construction of regular solutions even within LEDs. This formulation preserves the Maxwell limit in the weak-field regime, a feature that is particularly rare in traditional NLED-based constructions, and offers a conceptually simpler and physically more consistent alternative for generating singularity-free spacetimes.

This work is structured as follows. In Section~\ref{sec2}, we present the GR field equations coupled to a scalar field and an interaction term between the scalar field and the NLED, considering static and spherical symmetry. Next, we discuss two fundamental questions about the formalism we use to obtain the solutions, both for the case with magnetic charges and for the case with electric charges. In Section~\ref{Model}, we analyze two common models for static and spherically symmetric BB geometries — Simpson–Visser and Bardeen spacetime — and treat the magnetic and electric cases separately. In Section~\ref{EC}, we briefly discuss the derivation and analysis of the energy conditions associated with these solutions. Finally, in Section~\ref{sec:concl}, we present a summary and concluding discussion of the results obtained. We will adopt the metric signature ($+,-,-,-$) and employ geometrized units with $G=1$ and $c=1$, which represent the gravitational constant and the velocity of light, respectively.

%%%%%%%%%%%%%%%%%%%%%%%%%%%%%%%%%%%%%%%%%%%%%%%%%%%%%%%%%%%%%%%% 
\section{Field equations coupled to nonlinear electrodynamics and scalar field}\label{sec2}
%%%%%%%%%%%%%%%%%%%%%%%%%%%%%%%%%%%%%%%%%%%%%%%%%%%%%%%%%%%%%%%%

\subsection{Action and field equations}

We start by considering an action comprising the Einstein-Hilbert gravitational term, a scalar field, and a nonminimal interaction term coupling the scalar and electromagnetic sectors. The action is expressed as
\begin{equation}
S = \int \sqrt{-g} \, d^4x \left[ R - 2\kappa^2\big( \mathcal{L}_\varphi(\varphi) - \mathcal{L} _I (\varphi,F)\big) \right], \label{action}
\end{equation}
where $\kappa^2 = 8\pi$, $\mathcal{L}_\varphi(\varphi)$ is the Lagrangian associated with the scalar field, and $\mathcal{L} _I(\varphi,F)$ represents the interaction term between the scalar field and the electromagnetic sector.

The Lagrangian of the scalar field is written as
\begin{equation}
\mathcal{L}_\varphi(\varphi) = \epsilon(\varphi)\partial^\mu\varphi \, \partial_\mu\varphi - V(\varphi),\label{Lphi}
\end{equation}
where $\epsilon(\varphi) > 0$ corresponds to the canonical scalar field, while $\epsilon(\varphi) < 0$ defines a phantom-type scalar field. The term $V(\varphi)$ represents the scalar field potential. 

In this work, we consider the following explicit form for the interaction term:
\begin{equation}
\mathcal{L}_I(\varphi,F) = W(\varphi)\,\mathcal{L}(F),\label{LI}
\end{equation}
where $\mathcal{L}(F)$ is the Lagrangian density of the NLED, which depends on the electromagnetic invariant $F = \frac{1}{4}F^{\mu\nu}F_{\mu\nu}$, and $W(\varphi)$ is a function that establishes the nonminimal coupling between the electromagnetic and scalar fields. Note that when we take $W(\varphi)=1$, we recover the Lagrangian of the usual NLED. From this point on, we will develop our solutions based on the action \eqref{action}, with the definitions provided by Eqs.~\eqref{Lphi} and~\eqref{LI}.

The electromagnetic tensor $F_{\mu\nu}$ is defined in terms of the vector potential $A_\mu$ as
\begin{equation}
F_{\mu\nu} = \partial_\mu A_\nu - \partial_\nu A_\mu .
\end{equation}

Now, varying the action \eqref{action} with respect to the potential $A_\mu$ and the field $\varphi$, we find the following equations of motion:
\begin{equation}
\nabla_\mu (W(\varphi){\cal L}_F (F)F^{\mu\nu})=0,\label{sol2}
\end{equation}
and
\begin{eqnarray}
    2\epsilon(\varphi)\nabla_\mu\nabla^\mu \varphi+\nabla_\mu\varphi\nabla^\mu \varphi\frac{d\epsilon(\varphi)}{d\varphi}
     \nonumber\\
     +{\cal L}(F)\frac{dW(\varphi)}{d\varphi}=-\frac{dV(\varphi)}{d\varphi}\,,
     	\label{sol3}
\end{eqnarray}
respectively, where we denote ${\cal L}_F=\partial {\cal L}  (F)/\partial F$.

The gravitational field equation is obtained by varying the action \eqref{action} with respect to the metric tensor $g_{\mu\nu}$, and is given by
\begin{equation}
G_{\phantom{\mu}\nu}^{\mu}\equiv R_{\phantom{\mu}\nu}^{\mu}-\frac{1}{2}\delta{}_{\phantom{\mu}\nu}^{\mu}R=\kappa^{2}{T}{}_{\phantom{\mu}\nu}^{\mu}=\kappa^{2}\left(W(\varphi)\,\overset{F}{T}{}_{\phantom{\mu}\nu}^{\mu}+\overset{\varphi}{T}{}_{\phantom{\mu}\nu}^{\mu}\right).\label{EqM}
\end{equation}

Furthermore, the NLED energy-momentum tensor is defined as
\begin{align}
    \overset{F}{T}{}_{\phantom{\mu}\nu}^{\mu}=\delta_{\phantom{\mu}\nu}^{\mu}{\cal L}_{{\rm NLED}}(F)-{\cal L}_{F}F^{\mu\alpha}F_{\nu\alpha}\,,\label{TNED}
\end{align}
and the energy-momentum tensor for the scalar field matter part is expressed as follows
\begin{equation}
    \overset{\varphi}{T}{}_{\phantom{\mu}\nu}^{\mu}=\epsilon\,\Big(2\,\partial^{\mu}\varphi\partial_{\nu}\varphi-\delta_{\phantom{\mu}\nu}^{\mu}\partial^{\sigma}\varphi\partial_{\sigma}\varphi\Big)+\delta_{\phantom{\mu}\nu}^{\mu}V(\varphi)\,.\label{Tphi}
\end{equation}

Next, we introduce the spherically symmetric metric employed in our analysis, along with the algebraic framework used to derive the solutions.

%%%%%%%%%%%%%%%%%%%%%%%%%%%%%%%%%%%%%%%%%%%%%%%%%
\subsection{Spherically symmetric black bounce}
%%%%%%%%%%%%%%%%%%%%%%%%%%%%%%%%%%%%%%%%%%%%%%%%%

To begin developing our BB solutions, we consider a static and spherically symmetric line element
\begin{align}
&ds^2=A(r)dt^2-B(r)dr^2
-\Sigma(r)^2 \Big(d\theta^{2}+\sin^{2}\theta\,d\phi^{2}\Big),\label{m}
\end{align}
where $A(r)$ and $B(r)$ are functions of the radial coordinate $r$, and $\Sigma(r)$ is a non-trivial function that will implement the bounce in the radial coordinate.

The non-trivial components of the Einstein tensor \eqref{EqM}, for the metric \eqref{m}, are the following:
\begin{eqnarray}    G_{\phantom{0}0}^{0}&=&\Big[\Sigma(r)B'(r)\Sigma'(r)-B(r)\left(2\Sigma(r)\Sigma''(r)+\Sigma'(r)^{2}\right)
    \nonumber \\
    &&+B(r)^{2}\Big]\big/B(r)^{2}\Sigma(r)^{2}\,,
    \nonumber
    \\
    G_{\phantom{0}1}^{1}&=&-\frac{\Sigma(r)A'(r)\Sigma'(r)+A(r)\left(\Sigma'(r)^{2}-B(r)\right)}{A(r)B(r)\Sigma(r)^{2}}\,,
    \nonumber
    \\
G_{\phantom{0}2}^{2}&=&\frac{B'(r)\left(\Sigma(r)A'(r)+2A(r)\Sigma'(r)\right)}{4A(r)B(r)^{2}\Sigma(r)}
	\nonumber \\
		&&
\hspace{-0.75cm} +\Big[-2A(r)\left(A'(r)\Sigma'(r)+2A(r)\Sigma''(r)\right)
	\nonumber \\
		&&
\hspace{-0.75cm}+\Sigma(r)\left(A'(r)^{2}-2A(r)A''(r)\right)\Big] \big/\left(4A(r)^{2}B(r)\Sigma(r)\right).
	\nonumber
\end{eqnarray}

The symmetry of the metric \eqref{m} allows us to consider only radial electric and/or magnetic fields. The only non-zero components of the Maxwell-Faraday tensor $F_{\mu\nu}$ are therefore $F_{01} = -F_{10}$ and/or $F_{23} = -F_{32}$, which correspond to the contributions of the electric and/or magnetic charges.

In this paper, we treat separately the purely magnetic ($q_m \neq 0$, $q_e = 0$) and the purely electric ($q_e \neq 0$, $q_m = 0$) case. In the magnetic case, the only relevant components of $F^{\mu\nu}$ are the following
\begin{equation}
F^{23} = \frac{q_m \csc\theta}{\Sigma^4(r)}, \label{23}
\end{equation}
where $q_m$ represents the magnetic charge of the monopole. With this, the electromagnetic scalar $F$ takes the form:
\begin{equation}
F = \frac{q_m^2}{2\Sigma^4(r)}. \label{F}
\end{equation}

For the purely electrical case, the non-zero component is:
\begin{equation}
F^{10} = \frac{q_e}{W(r)\mathcal{L}_F(r)\Sigma^2(r)}, \label{10}
\end{equation}
where $q_e$ is the electric charge. Thus, the scalar $F$ now takes the form:
\begin{equation}
F = -\frac{q_e^2 A(r)B(r)}{2\mathcal{L}_F^2(r)\Sigma^4(r)W^2(r)}. \label{Fe}
\end{equation}

In addition, we will use the following identity
\begin{equation}
\mathcal{L}_F - \frac{\partial \mathcal{L}}{\partial r} \left( \frac{\partial F}{\partial r} \right)^{-1} = 0, \label{RC}
\end{equation}
which will be useful for checking the consistency of the solutions obtained.

%%%%%%%%%%%%%%%%%%%%%%%%%%%%%%%%%%%%%%%%%%%%%%%%%%%%%%%%%%%%
\subsubsection{Magnetics solutions}
%%%%%%%%%%%%%%%%%%%%%%%%%%%%%%%%%%%%%%%%%%%%%%%%%%%%%%%%%%%%

In this case, we illustrate the general solutions by initially considering the Maxwell-Faraday tensor with only a magnetic charge component, given by Eq. \eqref{23}.

Substituting the metric \eqref{m} and Eqs. \eqref{23} and \eqref{F}, into the Einstein field equation \eqref{EqM}, we obtain the following components for the right-hand side of the equations of motion:
\begin{align}
G_{\phantom{0}0}^{0}=\kappa^{2}T_{\phantom{0}0}^{0}=&\kappa^{2}\left(\frac{\epsilon(r)\varphi'(r)^{2}}{B(r)}+W(r){\cal L}(r)+V(r)\right),\label{EqF00}
\end{align}
\begin{align}
G_{\phantom{0}1}^{1}=\kappa^{2}T_{\phantom{0}1}^{1}=&\kappa^{2}\left(-\frac{\epsilon(r)\varphi'(r)^{2}}{B(r)}+W(r){\cal L}(r)+V(r)\right),\label{EqF11}
\end{align}
\begin{align}
G_{\phantom{0}2}^{2}=	\kappa^{2}T_{\phantom{0}2}^{2}=&\kappa^{2}\Bigg[W(r)\left({\cal L}(r)-\frac{q_m^{2}{\cal L}_{F}(r)}{\Sigma(r)^{4}}\right)
	\nonumber
    \\
    &
    +
    \frac{\epsilon(r)\varphi'(r)^{2}}{B(r)}+V(r)
    \Bigg]\,.\label{EqF22}
\end{align}

From Eqs. \eqref{EqF00} and \eqref{EqF22}, general expressions for the Lagrangian of the NLED $\mathcal{L}(r)$ and its derivative $\mathcal{L}F(r)$ can be determined, which are given by
\begin{align}
   & {\cal L}(r) =\Big\{ B(r)^{2}\Big(1-\kappa^{2}\Sigma(r)^{2}V(r)\Big)
    \nonumber
    \\
    & -B(r)\Big[\Sigma'(r)^{2}+\Sigma(r)\Big(2\Sigma''(r)+\kappa^{2}\Sigma(r)\epsilon(r)\varphi'(r)^{2}\Big)\Big]
    \nonumber
    \\
    &    +\Sigma(r)B'(r)\Sigma'(r)\Big\}\Big/\left(\kappa^{2}W(r)\Sigma(r)^{2}B(r)^{2}\right)\, ,	\label{L_BB}
\end{align}
and
\begin{eqnarray}
&&    {\cal L}_{F}(r)=\frac{\Sigma(r)^{2}}{4\kappa^{2}q_m^{2}A(r)^{2}B(r)^{2}W(r)}\Big\{-B(r)\Sigma(r)^{2}A'(r)^{2}
   \nonumber
    \\
&& \qquad
+2A(r)^{2}\big[-2B(r)\left(\Sigma(r)\Sigma''(r)+\Sigma'(r)^{2}\right)+2B(r)^{2}
   \nonumber
    \\
&& \qquad
+\Sigma(r)B'(r)\Sigma'(r)\big]+A(r)\Sigma(r)\big[2B(r)\Sigma(r)A''(r)
   \nonumber
    \\
&& \qquad
+A'(r)\left(2B(r)\Sigma'(r)-\Sigma(r)B'(r)\right)\big]\Big\}\,	,\label{LF_BB}
\end{eqnarray}
respectively.

We note that Eqs. \eqref{L_BB} and \eqref{LF_BB} satisfy the components \eqref{EqF00} and \eqref{EqF22} of Einstein's equations. However, Eq. \eqref{EqF11} provides an additional equation  that can be used to determine $\epsilon(r)$:
\begin{eqnarray}
   && \frac{2 A(r) B(r) \Sigma ''(r)-\Sigma '(r) \left(B(r) A'(r)+A(r) B'(r)\right)}{A(r) B(r)^2 \Sigma (r)}
    \nonumber\\
   && \hspace{4cm}  +\frac{2 \kappa ^2 \epsilon (r) \varphi '(r)^2}{B(r)}=0 \,.\label{comp00}
\end{eqnarray}
Using expression \eqref{comp00}, we can determine the parameter $\epsilon(r)$, in its general form, as follows 
\begin{equation}
    \epsilon(r)=\frac{\Sigma '(r) \Big(B(r) A'(r)+A(r) B'(r)\Big)-2 A(r) B(r) \Sigma ''(r)}{2 \kappa ^2 A(r) B(r) \Sigma (r) \varphi '(r)^2}.\label{eps}
\end{equation}

To determine the potential $V(r)$, we use the equation of motion \eqref{sol3}, which after substituting $\epsilon(r)$, i.e., Eq. \eqref{eps}, and direct integration leads to the following general expression
\begin{widetext}
\begin{eqnarray} 
	V(r) &=& W(r)\int\frac{1}{2\kappa^{2}W(r)^{2}A(r)\Sigma(r)^{2}B(r)^{3}}\Bigg\{ B(r)^{2}\Big[\Sigma'(r)\Big[\Sigma(r)W(r)A''(r)+A'(r)\Big(\Sigma(r)W'(r)+3W(r)\Sigma'(r)\Big)\Big]\Big]
	\nonumber\\&&
	\qquad \quad -B(r)^{2}\Sigma(r)W(r)A'(r)\Sigma''(r)+A(r)\Bigg[-2\Sigma(r)W(r)B'(r)^{2}\Sigma'(r)+B(r)\Bigg(3\Sigma(r)W(r)B'(r)\Sigma''(r)
	\nonumber\\&&
	\qquad \quad  -2B(r)^{2}W'(r)
	+\Sigma'(r)\Big[\Sigma(r)W(r)B''(r)+B'(r)\Big(3W(r)\Sigma'(r)-\Sigma(r)W'(r)\Big)\Big]
	\nonumber\\&&
	\qquad \quad +2B(r)\Big[W'(r)\Big(\Sigma(r)\Sigma''(r)+\Sigma'(r)^{2}\Big)
	-W(r)\left(\Sigma(r)\Sigma'''(r)+3\Sigma'(r)\Sigma''(r)\Big)\right]\Bigg)\Bigg]\Bigg\}dr
	.\label{V}
\end{eqnarray}
\end{widetext}

With the expressions \eqref{L_BB}, \eqref{LF_BB}, \eqref{eps}, and \eqref{V}, we have all the relevant quantities for modeling a solution with a purely magnetic source.
In the next subsection, we analyze the complementary case in which we only consider the presence of electric charge.

%%%%%%%%%%%%%%%%%%%%%%%%%%%%%%%%%%%%%%%%%%%%%%%%%%%%%%%%%%%%
\subsubsection{ Electric solutions} %Case electric
%%%%%%%%%%%%%%%%%%%%%%%%%%%%%%%%%%%%%%%%%%%%%%%%%%%%%%%%%%%%

In this subsection, we follow the same approach as before, but consider only the presence of electric charge, as expressed in Eq. \eqref{10}. Therefore, the only relevant component of the Maxwell-Faraday tensor that is not zero is $F^{10} \neq 0$.

Therefore, by including the contribution of the electric charge from Eq.~\eqref{10} into the equations of motion \eqref{EqM}, we explicitly obtain the following expressions.
\begin{eqnarray}
G_{\phantom{0}0}^{0}=\kappa^{2}T_{\phantom{0}0}^{0}&=&\kappa^{2}\Bigg(\frac{q_{e}^{2}A(r)B(r)}{{\cal L}_{F}(r)\Sigma^{4}(r)W(r)}+\frac{\epsilon(r)\varphi'^{2}(r)}{B(r)}
\nonumber
    \\
    &&
    +{\cal L}(r)W(r)+V(r)\Big)
    \,,\label{EqF00e}
\end{eqnarray}
\begin{eqnarray}
G_{\phantom{0}1}^{1}=\kappa^{2}T_{\phantom{0}1}^{1}&=&\kappa^{2}\Bigg(\frac{q_{e}^{2}A(r)B(r)}{{\cal L}_{F}(r)\Sigma(r)^{4}W(r)}-\frac{\epsilon(r)\varphi'^{2}(r)}{B(r)}
\nonumber
    \\
    &&
    +{\cal L}(r)W(r)+V(r)\Big)
    \,,\label{EqF11e}
\end{eqnarray}
\begin{equation}
G_{\phantom{0}2}^{2}=	\kappa^{2}T_{\phantom{0}2}^{2}=\kappa^{2}\left(\frac{\epsilon(r)\varphi'^{2}(r)}{B(r)}+{\cal L}(r)W(r)+V(r)\right)
	.\label{EqF22e}
\end{equation}
We now solve Eqs. \eqref{EqF00e} and \eqref{EqF22e} to find the Lagrangian $\mathcal{L} (r)$ and its derivative, $\mathcal{L}_F (r)$. The general expression for the Lagrangian of the NLED, which only considers the electric charge, is
\begin{eqnarray}
    &&{\cal L}(r)=\bigg\{ A(r)\Big[A'(r)\Big(\Sigma(r)B'(r)-2B(r)\Sigma'(r)\Big)
    \nonumber\\&&
    -2B(r)\Sigma(r)A''(r)\Big]-2A(r)^{2}\Big\{-B'(r)\Sigma'(r)
\nonumber\\&&
+2B(r)\Big[\kappa^{2}\Sigma(r)\Big(B(r)V(r)+\epsilon(r)\varphi'(r)^{2}\Big)+\Sigma''(r)\Big]\Big\}
\nonumber\\&&
+B(r)\Sigma(r)A'(r)^{2}\bigg\}\Big/\Big(4\kappa^{2}A(r)^{2}B(r)^{2}\Sigma(r)W(r)\Big)
    \,.	\label{Le_BB} 
\end{eqnarray}
While the derivative $\mathcal{L}_F(r)$ takes the form
\begin{eqnarray}
    &&{\cal L}_F(r)=4\kappa^{2}q_{e}^{2}A(r)^{3}B(r)^{3}\big/\bigg\{\Sigma(r)^{2}W(r)\Big\{A(r)\Sigma(r)
    \nonumber\\&&    \times\Big[2B(r)\Sigma(r)A''(r)+A'(r)\Big(2B(r)\Sigma'(r)-\Sigma(r)B'(r)\Big)\Big]
\nonumber\\&&
-B(r)\Sigma(r)^{2}A'(r)^{2}+2A(r)^{2}\Big[\Sigma(r)B'(r)\Sigma'(r)+2B(r)^{2}
\nonumber\\&&
-2B(r)\left(\Sigma(r)\Sigma''(r)+\Sigma'(r)^{2}\right)\Big]\Big\}\bigg\}
    \,.	\label{LFe_BB} 
\end{eqnarray}

Something similar to the magnetic case occurs here. The two expressions above satisfy Eqs. \eqref{EqF00e} and \eqref{EqF22e}. However, if we substitute Eqs. \eqref{Le_BB} and \eqref{LFe_BB} into Eq. \eqref{EqF11e}, the resulting equation is identical to the one obtained previously, namely, Eq.~\eqref{comp00}, and the function $\epsilon(r)$ is therefore still given by \eqref{eps}.

However, we note considerable differences in the arbitrary expression for the potential $V(r)$, which taking into account the form for the electric case of Eq. \eqref{sol3}, results in
\begin{widetext}
\begin{eqnarray}
    V(r)&=&W(r)\int \Bigg\{ \bigg\{ \Big[6A(r)W(r)B'(r)\Sigma'(r)^{2}-\Sigma(r)^{2}A'(r)B'(r)W'(r)+2A(r)\Sigma(r)W(r)\Big(B''(r)\Sigma'(r)+3B'(r)\Sigma''(r)\Big)\Big]
    \nonumber\\&&
   \qquad \times A(r)B(r)
   +
    B(r)^{2}\Big\{2A(r)\Sigma(r)\Big[W(r)\Big(A''(r)\Sigma'(r)-2A(r)\Sigma'''(r)\Big)+A'(r)\Big(2\Sigma'(r)W'(r)-W(r)\Sigma''(r)\Big)\Big]
    \nonumber\\&&    +6A(r)W(r)\Sigma'(r)\Big(A'(r)\Sigma'(r)-2A(r)\Sigma''(r)\Big)-\Sigma(r)^{2}W'(r)\Big(A'(r)^{2}-2A(r)A''(r)\Big)\Big\}
    \nonumber\\&&
   -4A(r)^{2}\Sigma(r)W(r)B'(r)^{2}\Sigma'(r)\bigg\} \Big/\Big( 4 \kappa ^2 A(r)^2 B(r)^3 \Sigma (r)^2 W(r)^2\Big)\Bigg\}dr\,
.\label{Ve_BB}
\end{eqnarray}
%\end{widetext}

Thus, with the expressions we have obtained in this subsection, we complete the equations required for the general analysis of the system with nonminimal coupling between the scalar field and NLED. Starting from these expressions, it is sufficient to make a functional choice for the metric functions $A(r)$, $B(r)$, the area $\Sigma(r)$, and the scalar field $\varphi(r)$ to obtain explicit solutions.
In the next section, we present specific examples of solutions resulting from the choice of these functions.

We also use the Kretschmann scalar to determine the regularity of spacetime. This quantity is described by the components of the Riemann tensor, which is given by $K=R_{\rho\sigma\mu\nu}R^{\rho\sigma\mu\nu}$. Its explicit form for metric \eqref{m} is
%
%\begin{widetext}
\begin{eqnarray}
K(r)&=&\bigg\{ B(r)^{2}\Sigma(r)^{4}A'(r)^{4}+A(r)^{2}\left(8B(r)^{2}\Sigma(r)^{2}A'(r)^{2}\Sigma'(r)^{2}+\Sigma(r)^{4}\big(A'(r)B'(r)-2B(r)A''(r)\big)^{2}\right)
\nonumber\\&&
\qquad +2A(r)B(r)\Sigma(r)^{4}A'(r)^{2}\left(A'(r)B'(r)-2B(r)A''(r)\right)+8A(r)^{4}\Big[2B(r)^{4}-4B(r)^{3}\Sigma'(r)^{2}
\nonumber\\&&
\qquad +\Sigma(r)^{2}B'(r)^{2}\Sigma'(r)^{2}+2B(r)^{2}\left(2\Sigma(r)^{2}\Sigma''(r)^{2}+\Sigma'(r)^{4}\right)-4B(r)\Sigma(r)^{2}B'(r)\Sigma'(r)\Sigma''(r)\Big]\bigg\}
\,.\label{K}
\end{eqnarray}
\end{widetext}

With this last presentation, we now have all the tools we need to develop our specific solutions and analyze them. In the next topic, we will present the models we get when we assume certain functional forms for metric functions, for example.

%%%%%%%%%%%%%%%%%%%%%%%%%%%%%%%%%%%%%%%%%%%%%%%%%
\section{Specific Models}\label{Model}
%%%%%%%%%%%%%%%%%%%%%%%%%%%%%%%%%%%%%%%%%%%%%%%%%

Before presenting the models, we begin with some preliminary considerations. In particular, we assume the following symmetry for the two cases under analysis:
\begin{align}
    B(r)=\frac{1}{A(r)}.\label{simetry}
\end{align}
With this choice, we aim to simplify the derivation of the solutions as much as possible.

From Eq. \eqref{simetry}, Eq. \eqref{eps} now takes the form of
\begin{equation}
    \epsilon(r)=-\frac{\Sigma''(r)}{2\kappa^{2}\Sigma(r)\varphi'(r)^{2}}.
\end{equation}

Note that, given the choice \eqref{simetry}, the function $\epsilon(r)$ does not depend explicitly on the metric functions $A(r)$ and $B(r)$, but only on the choice of the area function $\Sigma(r)$. Therefore, we will model this quantity as a phantom field, that is, we will assume $\epsilon(r) = -1$ in order to determine who is  $\varphi(r)$.
Furthermore, to continue our analysis, we will assume the following area function 
\begin{align}
    \Sigma(r)=\sqrt{a^2+r^2},\label{Sigma}
\end{align}
where $a$ is a real parameter with the dimension of length. This parameter will be interpreted, as appropriate, as the magnetic charge ($a = q_m$) or the electric charge ($a = q_e$).

Assuming that the scalar field is imaginary, i.e., $\epsilon(r)=-1$ and the area function defined in \eqref{Sigma}, we thus have the equation
\begin{equation}
    1=\frac{a^2}{ \kappa ^2 \left(a^2+r^2\right)^2 \varphi '(r)^2},
\end{equation}
which is a solution already well known in the literature on BB spacetimes, given by
\begin{equation}
    \varphi(r) = \frac{1}{\kappa}\,\tan^{-1}\left(\frac{r}{a}\right)\label{phi}.
\end{equation}
The range of the scalar field, resulting from the asymptotic analysis, is given by $
 -\frac{\pi}{2 \kappa} < \varphi < \frac{\pi}{2 \kappa}$.

These choices, defined by the symmetry~\eqref{simetry}, the area function~\eqref{Sigma}, and the scalar field~\eqref{phi}, will serve as the foundation for the construction of the models that follow. We emphasize that these assumptions will be maintained in all cases:
\begin{numcases}{\Sigma(r) = }
  \sqrt{q_m^2+r^2}, \quad \text{if} \quad a = q_m, \label{sig_qm} \\
  \nonumber\\
  \sqrt{q_e^2+r^2}, \quad \text{if} \quad a = q_e, \label{sig_qe}
\end{numcases}
and
\begin{numcases}{\varphi(r)=}
  \frac{\tan^{-1}\left(\frac{r}{q_m}\right)}{\kappa}, \quad \text{if} \quad a = q_m ,\label{phi_qm} \\
  \nonumber\\
  \frac{\tan^{-1}\left(\frac{r}{q_e}\right)}{\kappa}, \quad \text{if} \quad a = q_e .\label{phi_qe}
\end{numcases}

With these considerations, we conclude the presentation of the mathematical formalism necessary for the development of the models. In the following subsections, we propose two specific forms for the metric function: the Simpson-Visser type solution and, then the Bardeen type solution. In both geometries, we will investigate the purely magnetic ($q_m \neq 0$, $q_e = 0$) and purely electric ($q_m = 0$, $q_e \neq 0$) cases.

%%%%%%%%%%%%%%%%%%%%%%%%%%%%%%%%%%%%%%%%%%%%%%%%%
\subsection{Simpson-Visser type model}
%%%%%%%%%%%%%%%%%%%%%%%%%%%%%%%%%%%%%%%%%%%%%%%%%

In this first model, we start from the metric function proposed by Simpson-Visser, given by 
\begin{align}
 A(r)&=1-\frac{2 M}{\sqrt{a^2+r^2}}.
 \label{AMod1}
\end{align}

%%%%%%%%%%%%%%%%%%%%%%%%%%%%%%%%%%%%%%%%%%%%%%%%%
\subsubsection{Simpson-Visser model with magnetic charge:  $q_m\neq 0$ and $q_e=0$}
%%%%%%%%%%%%%%%%%%%%%%%%%%%%%%%%%%%%%%%%%%%%%%%%%

We begin the construction of this model by interpreting the parameter $a$ exclusively as the magnetic charge $q_m$. Thus, in addition to the symmetry established in \eqref{simetry}, we employ the forms of the area function $\Sigma(r)$ and the scalar field $\varphi(r)$ given in \eqref{sig_qm} and \eqref{phi_qm}, respectively. Under these conditions, the metric function \eqref{AMod1} takes the form
\begin{align}
      A(r)&=1-\frac{2 M}{\sqrt{q_m^2+r^2}}.\label{AMod1qm}
     \end{align}

After the substitution of Eqs. \eqref{simetry}, \eqref{sig_qm}, \eqref{phi_qm} and \eqref{AMod1qm} into the Lagrangian \eqref{L_BB}, we observe that it is still not possible to write ${\cal L}(r)$ explicitly and analytically as a function of the radial coordinate. This is because $W(r)$ remains undefined, which prevents the direct resolution of the integral that appears in the expression
\begin{align}
    {\cal L}(r)=&\frac{2M}{\kappa^{2}}q_m^{2}\left[\int\frac{\left(q_m^{2}+r^{2}\right)W'(r)+2rW(r)}{\left(q_m^{2}+r^{2}\right)^{7/2}W(r)^{2}}\,dr\right.
    \nonumber
    \\&
    \left.+\frac{1}{\left(q_m^{2}+r^{2}\right)^{5/2}W(r)}\right]\,.
\end{align}

To overcome this difficulty, we propose a simplification: We model the derivative of the Lagrangian, ${\cal L}_{F}(r)$, according to Eq.~\eqref{LF_BB}, and assume that it obeys a power law. This choice allows us to solve the differential equation and determine a functional form for $W(r)$. More precisely, we assume
\begin{equation}
    {\cal L}_F(r)=\alpha F^n,\label{Pot}
\end{equation}
where $\alpha$ is a constant, $n$ a free parameter, and $F$ the electromagnetic scalar.
With this assumption, it is possible to determine a solution for $W(r)$. Now, inserting Eq. \eqref{Pot} into Eq.~\eqref{LF_BB}, we obtain 
\begin{equation}
   \frac{3M}{\kappa^{2}W(r)\sqrt{q_m^{2}+r^{2}}}=\alpha  F^n,
\end{equation}
and after substituting Eq. \eqref{F} provides the following:
\begin{equation}
W(r)=\frac{3M2^{n}q_m^{-2n}\left(q_m^{2}+r^{2}\right)^{2n-\frac{1}{2}}}{\alpha \kappa^{2}}.
   \label{W}
\end{equation}

Thus, now with the solution \eqref{W}, we now express the Lagrangian \eqref{L_BB} in the following form
\begin{equation}
    {\cal L}(r)=\frac{2^{-n-1}\alpha \,q_m^{2n+2}\left(q_m^{2}+r^{2}\right)^{-2(n+1)}}{n+1},\label{L1}
\end{equation}
while its derivative \eqref{LF_BB} becomes
\begin{equation}
    {\cal L}_F(r)=2^{-n} \alpha  \, q_m^{2 n} \left(q_m^2+r^2\right)^{-2 n}.
\end{equation}

The combination of Eqs.\;\eqref{W} and \eqref{L1} provides the coupling term, which is written directly as a function of the radial coordinate:
\begin{equation}
   W(r){\cal L}(r) =\frac{3 M q_m^2}{2 \kappa ^2 (n+1) \left(q_m^2+r^2\right)^{5/2}}.
\end{equation}

From the expression \eqref{F}, we can write $r(F)$, and obtain the Lagrangian with respect to the electromagnetic scalar field. In addition, we also represent the Lagrangian obtained by taking $n=-1$ directly from Eq. \eqref{Pot}, 
\begin{numcases}{ {\cal L}_{}(F)=}
  \alpha \frac{F^{n+1}}{n+1}, \quad n \in \mathbb{Z}, n \neq -1 ,\label{L_VS} \\
  \nonumber\\
  \frac{\alpha}{3}\left[4+\ln(8)-6\ln\left(\frac{q_{m}}{\sqrt{F}}\right)\right],\; n  =-1 . 
\end{numcases}

It is important to note that, by setting $n=0$ and $\alpha=1$ in Eq. \eqref{L_VS}, we recover Maxwell's Lagrangian. This implies that it is possible to construct  BB solutions even in a linear electrodynamics (LED) scenario — a highly relevant result, since traditionally regular black hole models, and also BB, require NLED sources to eliminate singularities. Recently, it was found that BB solutions can be generated when considering an LED \cite{Junior:2025sjr}. We emphasize that, in this and the following models we will only focus on solutions with integer $n$, i.e., Eq. \eqref{L_VS}, since the solutions with NLED have already been extensively explored. For this reason, we will focus on the case with $n$, in particular with $n = 0$.

Furthermore, we note that if we choose $n\to 1/4$ and $\alpha \to \dfrac{3\sqrt[4]{2}M}{\kappa ^2\sqrt{|q_m|}}$, in Lagrangian \eqref{L_VS}, we recover exactly the Lagrangian presented in \cite{Rodrigues2023}, Eq.~(35): 
 \begin{equation*}
     {\cal L}(F)=\frac{12\sqrt[4]{2}MF^{5/4}}{5\kappa^{2}\sqrt{|q_{m}|}}.
 \end{equation*}

With the function $W(r)$ given by Eq. \eqref{W1}, it is now also possible to express the potential $V(r)$ \eqref{V}, within the case where $n\neq -1$, for this model
\begin{equation}
    V (r)=\frac{M (4 n+1) q_m^2}{2 \kappa ^2 (n+1) \left(q_m^2+r^2\right)^{5/2}} \,.
\end{equation}

\begin{figure}[htb!]
	\centering
	\includegraphics[scale=0.55]{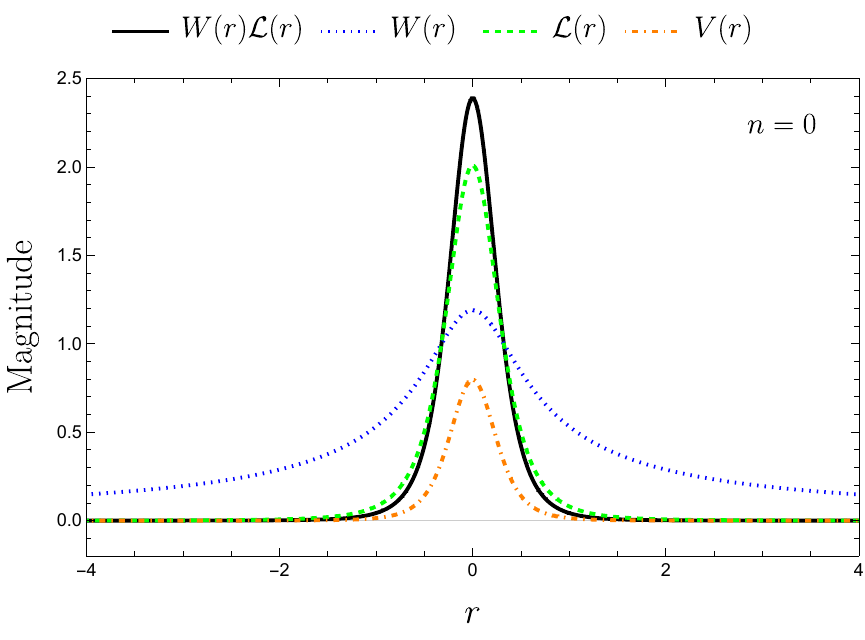} 
	\caption{Radial profiles of the functions provided in Eqs. \eqref{Eqn0} for $n=0$. The effective coupling term $W(r)\mathcal{L}(r)$ (solid black line), the individual functions $W(r)$ (dotted blue line) and $\mathcal{L}(r)$ (dashed green line), and the scalar potential $V(r)$ (orange dashed line).}
	\label{fig}
\end{figure}

In Fig. \ref{fig}, we present the radial profiles of the four main functions for the specific case of $n=0$: the coupling term $W(r)\mathcal{L}(r)$, the individual functions $W(r)$ and $\mathcal{L}(r)$, and the scalar potential $V(r)$. More specifically, we obtain under this condition the following relations:
\begin{align}
    W(r)=&\frac{3M}{\alpha\kappa^{2}\sqrt{q_{m}^{2}+r^{2}}},
    \nonumber\\
    W(r){\cal L}(r)=&\frac{3Mq_{m}^{2}}{2\kappa^{2}\left(q_{m}^{2}+r^{2}\right)^{5/2}},
    \nonumber\\
    {\cal L}(r)=&\frac{\alpha q_{m}^{2}}{2\left(q_{m}^{2}+r^{2}\right)^{2}},
    \nonumber\\
    V(r)=&\frac{Mq_{m}^{2}}{2\kappa^{2}\left(q_{m}^{2}+r^{2}\right)^{5/2}}
    \,, \label{Eqn0}
\end{align}
respectively.
To generate this plot, we used the following values for the constants: $\alpha \to \frac{3 \sqrt[4]{2} M}{\kappa ^2 \sqrt{|q_m|}}$, $M=5$, $q_m=0.5$ and $\kappa=\sqrt{8\pi}$.

The curve corresponding to $W(r)\mathcal{L}(r)$ (solid black line) directly represents the coupling term that is present in the action, and controls the intensity of the interaction between the scalar field and the electromagnetic sector. We expect the interaction term to correspond to the usual Lagrangian term in a BB description \cite{Rodrigues2023}, i.e., if $W(r)\mathcal{L}(r)\to\mathcal{L}(r)$, this term starts with a finite and positive value at the centre of the radial coordinate, and tends asymptotically to zero at large distances from $r$.

The individual profiles of $W(r)$ (blue dotted line) and $\mathcal{L}(r)$ (green dashed line) illustrate the different contributions of the two functions. Although both decrease with $r$, the function $\mathcal{L}(r)$ shows a steeper decline near the origin, reflecting the concentration of NLED effects in the central region.

Finally, the scalar potential $V(r)$ (orange dashed line) remains positive and follows a decay trend like the other functions, but with a smaller overall magnitude. Its profile confirms that the scalar field only makes a significant contribution near the central region, and becomes negligible at large distances.

This behavior confirms that the terms associated with the scalar field and NLED are responsible for the regularization of the central region, keeping the metric asymptotically flat and ensuring the recovery of general relativity at large scales.

%%%%%%%%%%%%%%%%%%%%%%%%%%%%%%%%%%%%%%%%%%%%%%%%%%%%%%%%%%

\begin{figure}[htb!]
\centering
\includegraphics[scale=0.4]{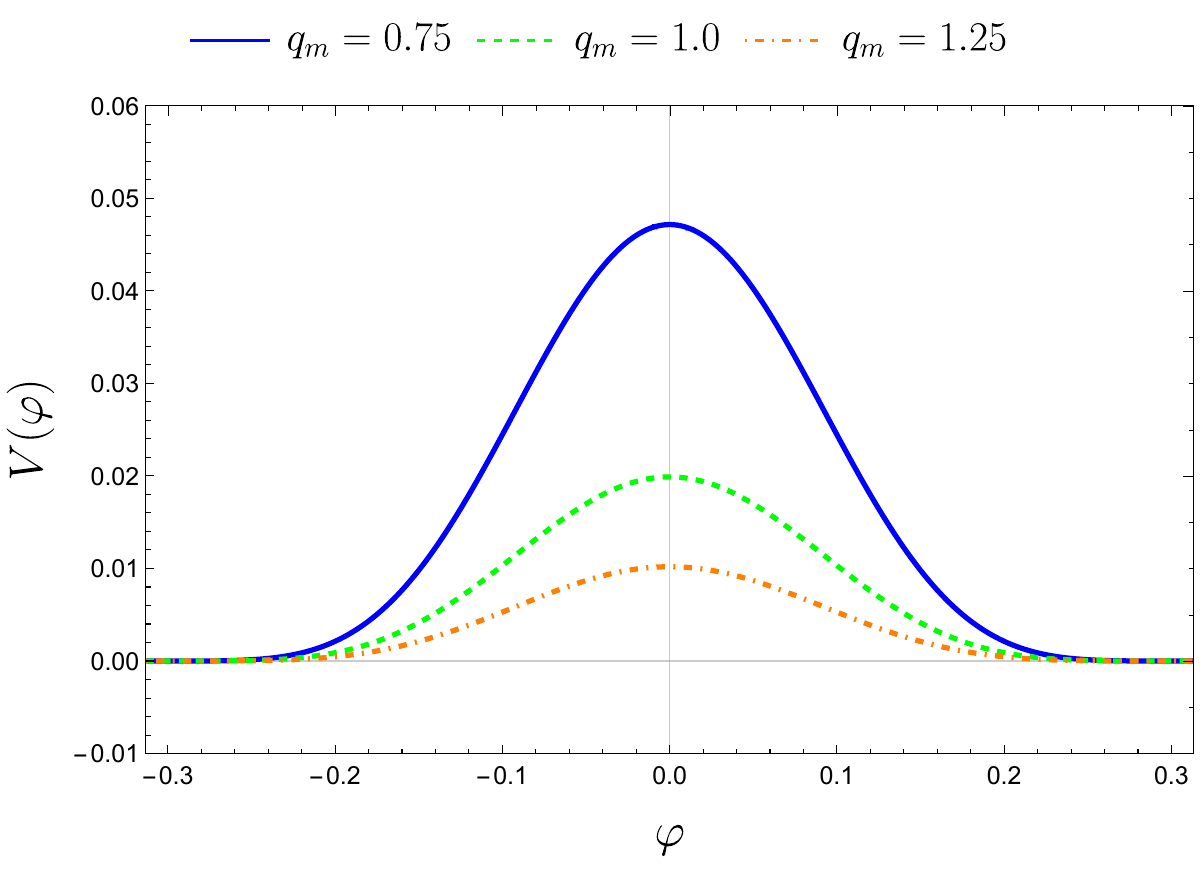} 
\caption{Potential behavior, as described by Eq. \eqref{V1}, for three different values of the magnetic charge at $n=0$ with $M=1$ and $\kappa=\sqrt{8\pi}$. }
\label{Vq1}
\end{figure}

Similar to the procedure applied to the Lagrangian ${\cal L}(F)$, we can write from the expression of the field \eqref{phi}, $r(\varphi)$, which allows us to write the potential directly as a function of the scalar field
\begin{equation}
    V (\varphi)=\frac{q_m^2 (4 M n+M)}{2 \kappa ^2 (n+1) | q_m \sec (  \kappa\varphi)| ^5} \,.\label{V1}
\end{equation}
Note also here that if we insert $n=1/4$ in equation \eqref{V1}, we get exactly the potential shown in equation (34) of the work \cite{Rodrigues2023}. In Fig. \ref{Vq1}, we illustrate the behavior of this potential for three different values of the magnetic charge if we consider $n=0$ with $M=1$ and $\kappa=\sqrt{8\pi}$.

%%%%%%%%%%%%%%%%%%%%%%%%%%%%%%%%%%%%%%%%%%%%%%%%%%%%%%%%%%

%%%%%%%%%%%%%%%%%%%%%%%%%%%%%%%%%%%%%%%%%%%%%%%%%%%%%%%%%%

Following the procedure for determining the potential $V (\varphi)$ Eq. \eqref{V1}, we can write the function \eqref{W} explicitly in terms of the scalar field. In this case, we obtain that
\begin{equation}
    W (\varphi)=\frac{3 M 2^n q_m^{-2 n} | q_m \sec ( \kappa\varphi )| ^{4 n-1}}{\alpha \kappa ^2} \,.\label{W1}
\end{equation}

Note that $W (\varphi)$ is not only related to the scalar field, but is also mixed with the gravitational sector, since in addition to $q_m$ it also depends on $M$, as shown in Eq. \eqref{W1}. In particular, larger values of $M$ increase the amplitude of $W(\varphi)$. Moreover, for $n=1/4$ and $\alpha = \dfrac{3\sqrt [4]{2}M}{\kappa^2\sqrt{|q_m|}}$, we obtain $W(\varphi)=1$, which corresponds to the case of the Simpson–Visser model without coupling between the scalar field and the electromagnetic sector, which is described only by the magnetic charge. Exactly as expected.

Therefore, using Eqs. \eqref{L_VS} and \eqref{W1}, we write the term representing the interaction Lagrangian in the action explicitly as
\begin{equation}
W(\varphi)\,\mathcal{L}(F)=\frac{3 M 2^n F^{n+1} q_m^{-2 n} \left| q_m \sec \left( \kappa \varphi \right)\right|^{4n-1}}{\kappa^2 (n+1)}\,.
\label{LISV}
\end{equation}
As already mentioned in the expressions for the radial coordinate $r$, the parameter $n$ determines the asymptotic behavior of the functions involved. In the interaction term $W(\varphi)\mathcal{L}(F)$, the dependence on $n$ follows the same principle, now expressed directly in terms of the scalar field $\varphi$ and the electromagnetic field $F$.

For the specific case of $n=0$, Eq.~\eqref{LISV} is reduced to the following simplified expression
\begin{equation}
W(\varphi)\,\mathcal{L}(F)=\frac{3 F M}{\kappa ^2 \sqrt{q_m^2 \sec ^2(\kappa  \varphi)}} \,,
\label{WL0}
\end{equation}
which exhibits a linear dependence on the electromagnetic field $F$.

\begin{figure}[htb!]
\centering
\includegraphics[scale=0.4]{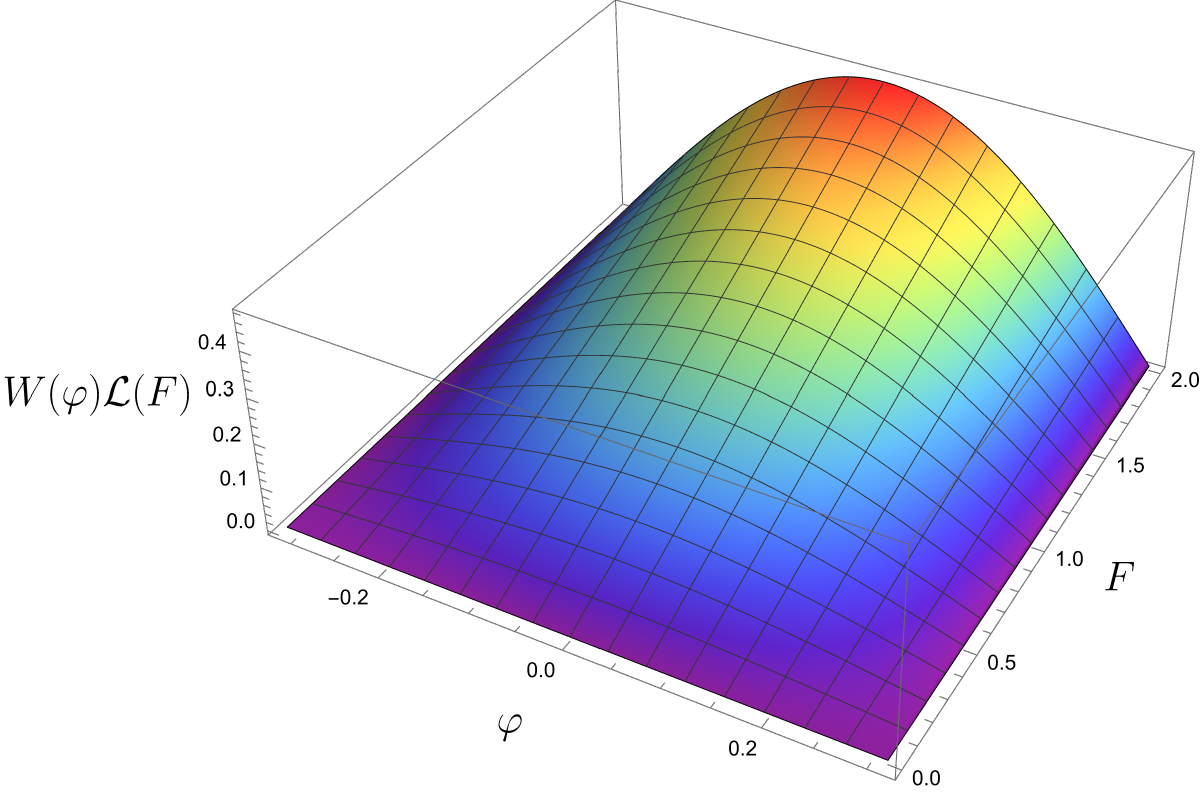} 
\caption{Behavior of the interaction term $W(\varphi)\,\mathcal{L}(F)$ for the case $n=0$, as given by Eq. \eqref{WL0}, with $q_m=0.5$ and $M=1$. The representation intervals are based on the asymptotic limits of the solution $\varphi(r)$, which lead to $-\frac{\pi}{2\kappa} < \varphi < \frac{\pi}{2\kappa}$, and on the maximum value of $F(r)$ at $r=0$, with $0 \leq F \leq \frac{1}{2q_m^2}$.}
\label{figLI}
\end{figure}

To illustrate the behavior of the interaction term described by Eq.~\eqref{WL0}, we present in Fig.~\ref{figLI} the three-dimensional graph corresponding to the case of $n=0$. The numerical parameters adopted have the values $q=0.5$ and $M=1$. 
The plotting intervals were defined as follows: $ -\frac{\pi}{2 \kappa} < \varphi < \frac{\pi}{2 \kappa}$  and for $F$  the interval used is $0\leq F \leq \frac{1}{2q_m^2}$, where the upper limit corresponds to the maximum value of $F(r)$ at the center of the configuration ($r=0$), as obtained by Eq.~\eqref{F}.

In the graph, the $F$-axis represents the electromagnetic contribution, while the $\varphi$-axis shows the influence of the scalar field on the modulation of the coupling term. For $n=0$, the dependence on $F$ is purely linear, but the modulation in $\varphi$ follows a dependence proportional to $|\sec( \kappa \varphi)|^{-1}$. As a result, a suppression of the interaction term is observed near the asymptotic limits of $\varphi$.

%%%%%%%%%%%%%%%%%%%%%%%%%%%%%%%%%%%%%%%%%%%%%%%%%%%%%%%%%%

%%%%%%%%%%%%%%%%%%%%%%%%%%%%%%%%%%%%%%%%%%%%%%%%%
\subsubsection{Simpson-Visser model with electric charge: $q_m=0$ and $q_e\neq0$}
%%%%%%%%%%%%%%%%%%%%%%%%%%%%%%%%%%%%%%%%%%%%%%%%%

We now develop the Simpson-Visser BB solution considering the electric charge $q_e$ as the source. To maintain this agreement, i.e., $a=q_e$, we will use the area function and the scalar field described in Eqs. \eqref{sig_qe} and \eqref{phi_qe}. This leads us to write Eq. \eqref{AMod1} as
\begin{align}
     A(r)&=1-\frac{2 M}{\sqrt{q_e^2+r^2}}.\label{AMod1qe}
     \end{align}

Under these circumstances, the electromagnetic scalar \eqref{Fe} has the form
\begin{align}
    F=-\frac{9 M^2 q_e^2}{2 \kappa ^4 \left(q_e^2+r^2\right)^3}.\label{Fe_SV}
\end{align}
Note that this invariant $F(r)$ generated by a regulated electric field is strictly negative in the context of BB-type solutions, i.e., $F(r) < 0$ for all $r \in (-\infty, \infty)$.

Analogous to the magnetic case, if we substitute the quantities obtained so far, in particular, Eqs.~\eqref{sig_qe}, \eqref{phi_qe} and \eqref{AMod1qe}, for example, into the Lagrangian \eqref{Le_BB}, we still cannot express $\mathcal{L}(r)$ analytically, since the function $W(r)$ is not defined. Thus, we obtain
\begin{align}
    {\cal L}(r)=&-\frac{Mq_{e}^{2}}{\kappa^{2}}\Bigg[\int\frac{\left(q_{e}^{2}+r^{2}\right)W'(r)-4rW(r)}{\left(q_{e}^{2}+r^{2}\right)^{7/2}W(r)^{2}}\,dr
   \nonumber \\&
    +\frac{1}{W(r)\left(q_{e}^{2}+r^{2}\right)^{5/2}}\Bigg].
\end{align}

To circumvent this difficulty, we proceed in the same way as in the magnetic case, i.e., we model the derivative of the Lagrangian ${\cal L}_{F}(r)$ as the power law, given by Eq.~\eqref{Pot}, but now apply it to Eq.~\eqref{LFe_BB} to determine $W(r)$.
The result is
\begin{equation}
   \frac{\kappa ^2 \sqrt{q_e^2+r^2}}{3 M W(r)}=\alpha F^n,
\end{equation}
and now we get the following solution, for $W(r)$
\begin{equation}
   W(r)=\frac{(-2)^n \kappa ^{4 n+2} \left(q_e^2+r^2\right)^{3 n+\frac{1}{2}}}{  3^{2 n+1} \alpha M^{2 n+1} q_e^{2 n}}.
   \label{We_SV}
\end{equation}

The solution \eqref{We_SV} provides us with the Lagrangian for this model with the following representation
\begin{equation}
    {\cal L}(r)=-\frac{(-1)^n\alpha  \left(\frac{2}{9}\right)^{-n-1}  M^{2 n+2} q_e^{2 n+2}}{(n+1) \kappa ^{4 n+4} \left(q_e^2+r^2\right)^{3 n+3}},\label{Le_SV}
\end{equation}
while its derivative \eqref{LFe_BB} is
\begin{equation}
    {\cal L}_F(r)=\frac{\alpha  \left(-\frac{2}{9}\right)^{-n} M^{2 n} q_e^{2 n}}{\kappa ^{4 n} \left(q_e^2+r^2\right)^{3 n}}.
\end{equation}

Thus, the interaction term, resulting from the Eqs. \eqref{We_SV} and \eqref{Le_SV}, is simply described by
\begin{equation}
   W(r){\cal L}(r) =-\frac{3 M q_e^2}{2 \kappa ^2 (n+1) \left(q_e^2+r^2\right)^{5/2}}, \label{WLe}
\end{equation}
which is regular for small values of $r$ and zero for large values of $r$. 

From Eq.~\eqref{Fe_SV}, the radial coordinate $r$ can be rewritten as a function of the electromagnetic invariant $F$, i.e. $r(F)$, and thus the Lagrangian \eqref{Le_SV} can be expressed directly in terms of $F$, which for this electrical case is
\begin{numcases}{ {\cal L}_{}(F)=}
  \frac{\alpha \,  F^{n+1}}{n+1}, \qquad n \in \mathbb{Z},\qquad n \neq -1 ,\label{Le2_SV} \\
  \nonumber\\
  -\frac{\alpha}{3}\Bigg[-2+9\ln\left(\frac{M^{2/3}q_{e}^{2/3}}{\sqrt[3]{|F|}\kappa^{4/3}}\right)
  	\nonumber\\
  \nonumber\\
 \phantom{-\frac{\alpha}{3}\Bigg[} 
 + \ln\left(\frac{729}{8}\right)\Bigg],\qquad n  =-1 . 
\end{numcases}

If we also use $n=-1/6$ and $\alpha=\left( \frac{|q e|^5 \kappa^4}{2^{1/2} \cdot 3 \cdot a^4 M^2} \right)^{1/3}$ in Eq. \eqref{Le2_SV}, we recover expression (24) presented in the article by BB~\cite{Alencar:2024yvh}. Next, we derive the relevant equations for the case where $n\neq -1$.

Using Eq.\;\eqref{We_SV}, for example, the potential \eqref{Ve_BB} now has the form
\begin{equation}
    V (r)=-\frac{M (2 n-1) q_e^2}{2 \kappa ^2 (n+1) \left(q_e^2+r^2\right)^{5/2}} \,.\label{Ve_SV}
\end{equation}

We can also write $r(\varphi)$, from the expression of the field \eqref{phi_qe},  which results in the following  potential 
\begin{equation}
    V (\varphi)=\frac{q_e^2 (M-2 M n)}{2 \kappa ^2 (n+1) \left(q_e^2 \sec ^2\left( \kappa  \varphi\right)\right)^{5/2}} \,.\label{Ve2_SV}
\end{equation}
Note that if we consider $n=-1/6$, we recover the same potential shown in Eq. (34) of the article~\cite{Alencar:2024yvh}. We also note that if we use $n=0$ in Eq. \eqref{Ve2_SV}, we find the same expression described in the magnetic case. For this reason, we do not present the graphical behavior again, since Fig.~\ref{Vq1} already adequately illustrates the corresponding profile.

\begin{figure}[htb!]
\centering
\includegraphics[scale=0.55]{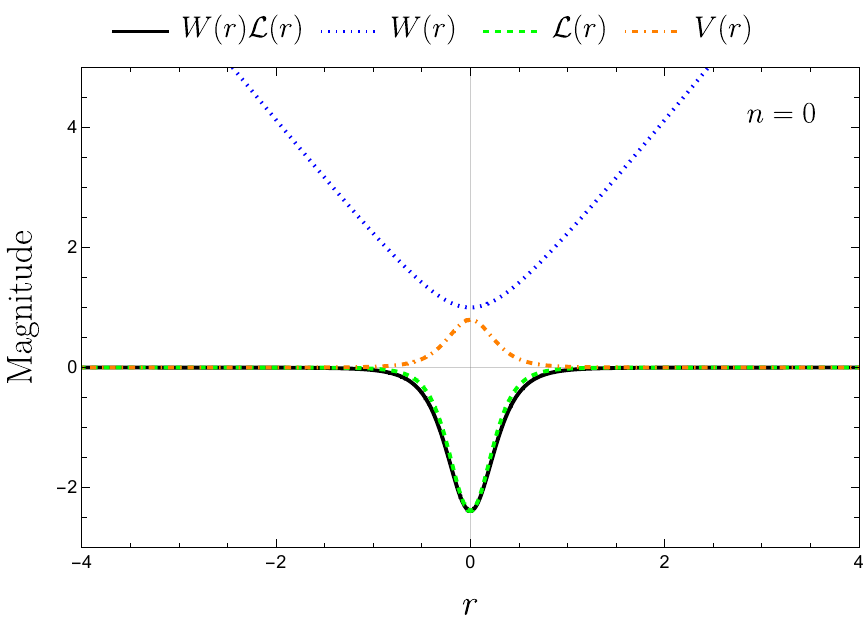} 
\caption{Radial profiles of the functions for $n=0$. The coupling term $W(r)\mathcal{L}(r)$ (solid black line), the individual functions $W(r)$ (dotted blue line) and $\mathcal{L}(r)$ (dashed green line), and the scalar potential $V(r)$ (orange dashed line), as described by Eqs. \eqref{Eqn0qe}, can be described. Parameters: $M=5$, $q_e=0.5$, $\kappa=\sqrt{8\pi}$. }
\label{fige}
\end{figure}

We also illustrate the behavior of the individual functions $W(r)$, given by Eq. \eqref{We_SV}, and $\mathcal{L}(r)$, i.e., Eq. \eqref{Le_SV}, the coupling term $W(r)\mathcal{L}(r)$, Eq. \eqref{WLe}, and the potential $V(r)$, provided by Eq. \eqref{Ve_SV}, for the special case of $n=0$, as shown in Fig. \ref{fige}. The explicit expressions with this consideration are:
\begin{align}    W(r)=&\frac{\kappa^{2}\sqrt{q_{e}^{2}+r^{2}}}{3\alpha M},
    \nonumber\\
    W(r){\cal L}(r)=&-\frac{3Mq_{e}^{2}}{2\kappa^{2}\left(q_{e}^{2}+r^{2}\right)^{5/2}},
    \nonumber\\
    {\cal L}(r)=&-\frac{9\alpha M^2 q_{e}^{2}}{2\kappa ^4\left(q_{e}^{2}+r^{2}\right)^{3}},
    \nonumber\\
    V(r)=&\frac{Mq_{e}^{2}}{2\kappa^{2}\left(q_{e}^{2}+r^{2}\right)^{5/2}}
    \, . \label{Eqn0qe}
\end{align}
To create this plot, we used the following values for the constants: $\alpha = \frac{\kappa ^2 |q_e|}{3 M}$, $M=5$, $q_e=0.5$ and $\kappa=\sqrt{8\pi}$.

The solid black curve represents $W(r)\mathcal{L}(r)$, while the dashed green curve corresponds to $\mathcal{L}(r)$. Both start with negative and finite values at $r=0$ and decay to zero as $r$ increases. The blue dotted curve shows the behavior of $W(r)$, which grows indefinitely with $r$, and the orange dashed-dotted curve describes $V(r)$, which is positive and tends towards zero at large distances.

%%%%%%%%%%%%%%%%%%%%%%%%%%%%%%%%%%%%%%%%%%%%%%%%%%%%%%%%%%

%%%%%%%%%%%%%%%%%%%%%%%%%%%%%%%%%%%%%%%%%%%%%%%%%%%%%%%%%%

%%%%%%%%%%%%%%%%%%%%%%%%%%%%%%%%%%%%%%%%%%%%%%%%%%%%%%%%%%

Analogous to the process performed to determine the potential $V (\varphi)$, given by Eq.~\eqref{Ve2_SV}, we can also express the function \eqref{We_SV} explicitly in the form of the scalar field. Thus, $W (\varphi)$ is now given by:
\begin{equation}
    W (\varphi)=\frac{(-2)^{n}\kappa^{4n+2}q_{e}^{-2n}\left(q_{e}^{2}\sec^{2}(\kappa\varphi)\right)^{3n+\frac{1}{2}}}{3^{1+2n1}M^{1+2n}\alpha} \,.\label{We2}
\end{equation}
Note that, if we consider in Eq.~\eqref{We2}, the values $n=-1/6$ and $\alpha=\left( \frac{|q e|^5 \kappa^4}{2^{1/2} \cdot 3 \cdot a^4 M^2} \right)^{1/3}$, we get $W(\varphi)=1$. As expected, in this limit, the model returns to the Simpson-Visser case with a purely electric source without additional coupling.

Thus, Eqs.~\eqref{Le2_SV} and \eqref{We2} explicitly provide us with the nonminimal coupling term as
\begin{equation}
W(\varphi)\,\mathcal{L}(F)=\frac{(-2)^{n}\,\,F^{n+1}\,\kappa^{4n+2}\,q_{e}^{4n+1}\,\sec^{6n+1}(\kappa\varphi)}{3^{2n+1}\,M^{2n+1}\,(n+1)} \,.
\label{LIe_SV}
\end{equation}

The interaction term Eq. \eqref{LIe_SV} reduces to the following result when we consider $n=0$
\begin{equation}
W(\varphi)\,\mathcal{L}(F)=\frac{\kappa^{2}F\sqrt{q_{e}^{2}\sec^{2}(\kappa\varphi)}}{3M} \,.
\label{LIe2_SV}
\end{equation}

\begin{figure}[htb!]
\centering
\includegraphics[scale=0.35]{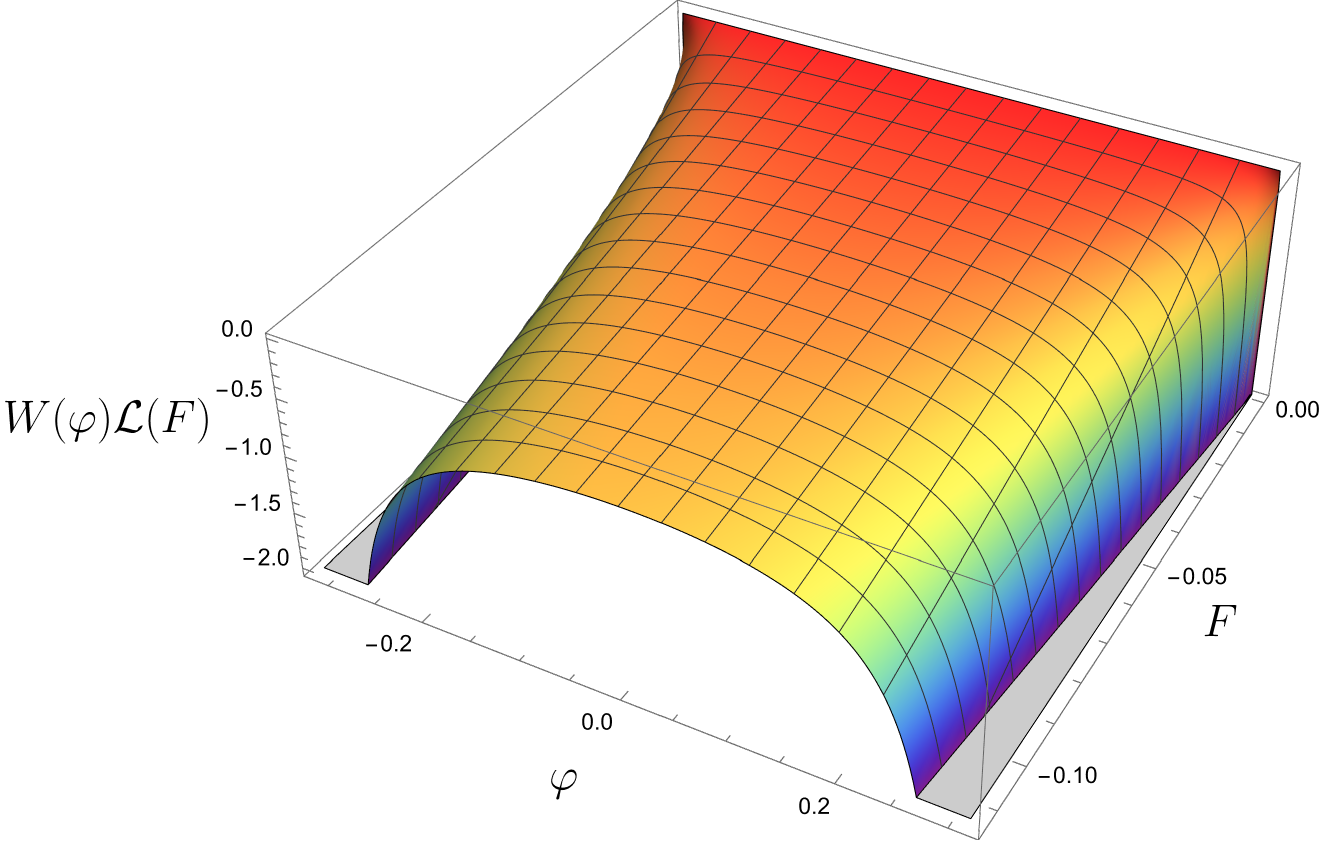} 
\caption{Behavior of the interaction term $W(\varphi)\,\mathcal{L}(F)$ for $n=0$, as described by Eq.~\eqref{LIe2_SV}, with $q_e=0.5$ and $M=1$. The plotting intervals are based on the asymptotic limits of the solution $\varphi(r)$, resulting in $-\frac{\pi}{2\kappa} < \varphi < \frac{\pi}{2\kappa}$, and on the maximum value of $F(r)$ at $r=0$, with $-\frac{9 M^2}{2 \kappa^4 q_e^4} \leq F \leq 0$. }
\label{figLISVe}
\end{figure}

To illustrate the behavior of the interaction term described by Eq.~\eqref{LIe_SV}, we present in Fig.~\ref{figLISVe} the three-dimensional graph, developed in a similar way to that of Fig.~\ref{figLI}, but now corresponding to the case with electric charge with the values of the constants being $q_e = 0.5$ and $M = 1$.
The plotting intervals were defined as follows: the interval of $\varphi$ is the same as previously considered, while for the field $F$ the interval $-\frac{9 M^2}{2 \kappa^4 q_e^4} \leq F \leq 0$ was adopted. The lower limit of this interval corresponds to the minimum value assumed by $F(r)$ at the center of the configuration ($r = 0$), as defined from Eq.~\eqref{Fe_SV}.
%%%%%%%%%%%%%%%%%%%%%%%%%%%%%%%%%%%%%%%%%%%%%%%%%%%%%%%%%%

%%%%%%%%%%%%%%%%%%%%%%%%%%%%%%%%%%%%%%%%%%%%%%%%%%%%

%%%%%%%%%%%%%%%%%%%%%%%%%%%%%%%%%%%%%%%%%%%%%%%%%%%%%%%%%%

\subsubsection{Kretschmann scalar}

Finally, to verify the regularity of spacetime, we calculate the Kretschmann scalar for this model from Eq. \eqref{K}. For this model, it is given by:
\begin{align}
    K=&\frac{12a^{4}}{\left(a^{2}+r^{2}\right)^{4}}+\frac{12M^{2}\left(3a^{4}-4a^{2}r^{2}+4r^{4}\right)}{\left(a^{2}+r^{2}\right)^{5}}
    \nonumber\\
    &
    +\frac{32Ma^{2}\left(r^{2}-a^{2}\right)}{\left(a^{2}+r^{2}\right)^{9/2}}\, .\label{K1}
\end{align}

Take the limit of Eq. \eqref{K1} for small values of $r$, we obtain
\begin{equation}
\lim_{r\to0}K=\frac{4\left(9M^{2}a^{4}-8M\sqrt{a^{2}}a^{4}+3a^{6}\right)}{a^{10}},
\end{equation}
which only depends on the constants of the model. 

For $r\to\infty$, we have $K(r)\to0$, which confirms the regularity of the spacetime and is consistent with the typical characteristics of BB solutions.

%%%%%%%%%%%%%%%%%%%%%%%%%%%%%%%%%%%%%%%%%%%%%%%%%
% \section{Model II}\label{MdII}
%%%%%%%%%%%%%%%%%%%%%%%%%%%%%%%%%%%%%%%%%%%%%%%%%
%%%%%%%%%%%%%%%%%%%%%%%%%%%%%%%%%%%%%%%%%%%%%%%%%
\subsection{Bardeen type model}
%%%%%%%%%%%%%%%%%%%%%%%%%%%%%%%%%%%%%%%%%%%%%%%%%

In this model, we adopt the same assumptions that we presented at the beginning of the Sec.~\ref{Model}, but now consider the metric function associated with the Bardeen-type solution
\begin{align}
 A(r)&=1-\frac{2Mr^{2}}{\left(a^{2}+r^{2}\right)^{3/2}}.
 \label{AMod2}
\end{align}
We also analyze solutions in two different scenarios. The first is described with magnetic charge only ($q_m\neq 0$, $q_e= 0$), the second soley with electric charge ($q_m = 0$, $q_e \neq 0$).

%%%%%%%%%%%%%%%%%%%%%%%%%%%%%%%%%%%%%%%%%%%%%%%%%
\subsubsection{Bardeen model with magnetic charge: $q_m\neq0$ and $q_e=0$}
%%%%%%%%%%%%%%%%%%%%%%%%%%%%%%%%%%%%%%%%%%%%%%%%%

In this first approach, we develop solutions by first considering only the magnetic charge, $q_m \neq 0$, and $q_e = 0$.
In this context, the functions $\Sigma(r)$ and $\varphi(r)$ are given, by the Eqs.~\eqref{sig_qm} and \eqref{phi_qm}, and the metric function \eqref{AMod2} has the form
\begin{align}
 A(r)=1-\frac{2Mr^{2}}{\left(q_m^{2}+r^{2}\right)^{3/2}}.
 \label{AMod2qm}
\end{align}

Using the symmetry given by Eq. \eqref{simetry}, and substituting Eqs. \eqref{sig_qm}, \eqref{phi_qm} and \eqref{AMod2qm}, into the Lagrangian \eqref{L_BB}, we realize that we cannot express ${\cal L}(r)$ analytically without an explicit definition for $W(r)$. In this particular case, we find that
\begin{align}
    &{\cal L}(r)=\frac{2Mq_m^{2}}{\kappa^{2}}\Bigg[\frac{3r^{2}}{\left(q_m^{2}+r^{2}\right)^{7/2}W(r)}
    \nonumber
    \\
    &
    +
\int\frac{r\left[3r\left(q_m^{2}+r^{2}\right)W'(r)+2\left(r^{2}-2q_m^{2}\right)W(r)\right]}{\left(q_m^{2}+r^{2}\right)^{9/2}W(r)^{2}}\,dr\Bigg].	
\end{align}

Similar to the procedure used for the Simpson-Visser model, we assume that ${\cal L}_{F}(r)$ follows a power law, Eq.~\eqref{Pot} (but we will consider in this model that $\alpha=1$), which allows us to obtain $W(r)$ from Eq. \eqref{LF_BB}, which is written here as:
\begin{equation}
   \frac{M\left(13r^{2}-2q_m^{2}\right)}{\kappa^{2}\left(q_m^{2}+r^{2}\right)^{3/2}W(r)}=F^n,
\end{equation}
which provides
\begin{equation}
   W(r)=-\frac{M2^{n}q_m^{-2n}\left(2q_m^{2}-13r^{2}\right)\left(q_m^{2}+r^{2}\right)^{2n-\frac{3}{2}}}{\kappa^{2}}.
   \label{W2}
\end{equation}

Therefore, using expression \eqref{W2},  we can express the Lagrangian
\eqref{L_BB} as follows
\begin{equation}
    {\cal L}(r)=\frac{2^{-n-1}q_m^{2n+2}\left(q_m^{2}+r^{2}\right)^{-2(n+1)}}{n+1},
\end{equation}
and its derivative, for this Bardeen metric function, is given by
\begin{equation}
    {\cal L}_F(r)=2^{-n} q_m^{2 n} \left(q_m^2+r^2\right)^{-2 n}.
\end{equation}

Therefore, the interaction factor in terms of the coordinate $r$ is
\begin{equation}
  W(r)\,\mathcal{L}(r)  =\frac{M\left(13q_{m}^{2}r^{2}-2q_{m}^{4}\right)}{2\kappa^{2}(n+1)\left(q_{m}^{2}+r^{2}\right)^{7/2}}\,.
\end{equation}

From Eq. \eqref{F}, we can express the Lagrangian in terms of $F$, as
\begin{numcases}{ {\cal L}_{}(F)=}
   \frac{F^{n+1}}{n+1}, \quad n \neq -1 ,\label{L_Bard} \\
  \nonumber\\
  \frac{12}{13}+\log(2)-2\log\left(\frac{|q_{m}|}{\sqrt{F}}\right),\quad n  =-1 . 
\end{numcases}
The result presented by Eq. \eqref{L_Bard} shows the same functional structure found in the Simpson-Visser case, and  for the case where $n=0$, we recover a linear Lagrangian (Maxwell type), which makes it possible to obtain regular Bardeen solutions even with linear electrodynamics -- something not reported in the literature, where Bardeen solutions traditionally require NLED to eliminate the singularity. As we developed earlier, we will focus on the case where $n\neq -1$ to derive the next physical quantities.

Continuing, the potential $V(r)$, given by Eq. \eqref{V}, after substituting all necessary quantities, such as Eq. \eqref{W2}, for instance, is now expressed as
\begin{equation}
    V(r)=\frac{Mq_m^{2}\left[(12n-1)r^{2}+2q_m^{2}\right]}{2\kappa^{2}(n+1)\left(q_m^{2}+r^{2}\right)^{7/2}} \,.
\end{equation}

%%%%%%%%%%%%%%

For $n=0$, the relevant functions become:
\begin{align}
W(r)=&\frac{M\left(13r^{2}-2q_{m}^{2}\right)}{\kappa^{2}\left(q_{m}^{2}+r^{2}\right)^{3/2}},
    \nonumber\\
    W(r){\cal L}(r)=&\frac{M\left(13q_{m}^{2}r^{2}-2q_{m}^{4}\right)}{2\kappa^{2}\left(q_{m}^{2}+r^{2}\right)^{7/2}},
    \nonumber\\
    {\cal L}(r)=&\frac{q_{m}^{2}}{2\left(q_{m}^{2}+r^{2}\right)^{2}},
    \nonumber\\
    V(r)=&\frac{M\left(2q_{m}^{2}-r^{2}\right)q_{m}^{2}}{2\kappa^{2}\left(q_{m}^{2}+r^{2}\right)^{7/2}}
    \,, \label{Eqn0qe_Bard}
\end{align}
respectively. 

\begin{figure}[htb!]
\centering
\includegraphics[scale=0.55]{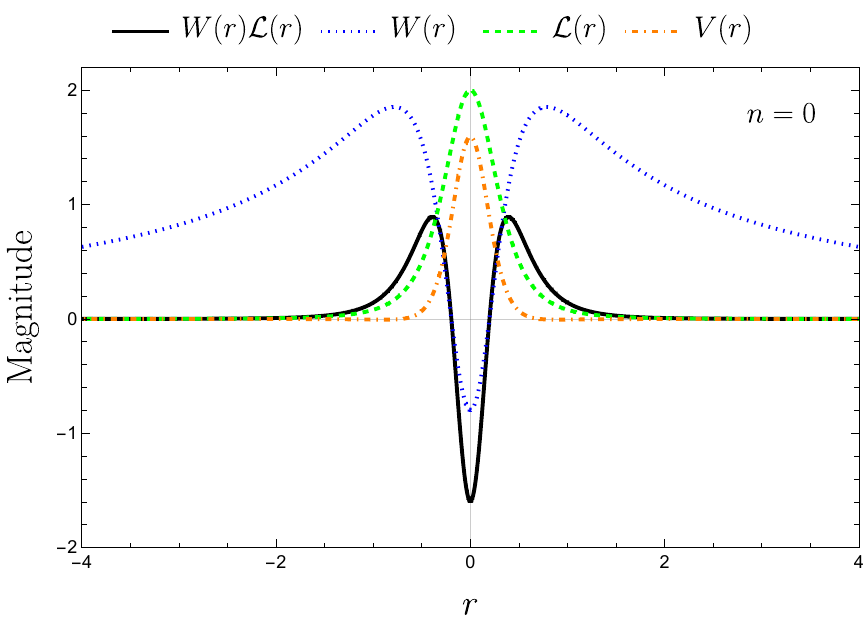} 
\caption{Behavior of the interaction term $W(\varphi)\,\mathcal{L}(F)$ for $n=0$, as described by Eq.~\eqref{Eqn0qe_Bard}, with $q_m=0.5$ and $M=1$. The plotting intervals are based on the asymptotic limits of the solution $\varphi(r)$, resulting in $-\frac{\pi}{2\kappa} < \varphi < \frac{\pi}{2\kappa}$, and on the maximum value of $F(r)$ at $r=0$, with $0 \leq F \leq \frac{1}{2q_m^2}$.}
% Perfis radiais para $n=0$: termo de acoplamento efetivo $W(r)\mathcal{L}(r)$ (linha preta sólida), funções individuais $W(r)$ (linha azul pontilhada) e ${\cal L}(r)$ (linha verde tracejada), e o potencial escalar $V(r)$ (linha laranja traço-pontilhada), conforme descritas pelas Eqs.~\eqref{Eqn0qe_Bard}, respectivamente. Parâmetros: $M=5$, $q_m=0.5$, $\kappa=\sqrt{8\pi}$.}
\label{figBardeenMag}
\end{figure}

Figure~\ref{figBardeenMag} illustrates the behavior of these functions. The dotted blue, dashed green, and dotted orange curves represent the quantities $W(r)$, $\mathcal{L}(r)$, and $V(r)$, respectively. The interaction term $W(r)\mathcal{L}(r)$ (solid black line) has a negative sign near $r=0$ and becomes positive at greater distances. This indicates a local transition between regimes: for $W(r)\mathcal{L}(r)>0$, the coupling describes a Lagrangian that acts as a canonical field, while for $W(r)\mathcal{L}(r)<0$, the Lagrangian takes on a phantom character. This fact indicates that we have a description of a partially phantom Lagrangian.
To create this graph, we used the following values for the constants: $M=5$, $q_m=0.5$ and $\kappa=\sqrt{8\pi}$.

%%%%%%%%%%%%%%%%%%%%%%%%%%%%%%%%%%%%%%%%%%%%%%%%%%%%%%%%%%

Writing $r(\varphi)$, we obtain the potential in terms of the scalar field, whose explicit form is found in the expression below
\begin{equation}
    V (\varphi)=\frac{M\cos^{4}\left(\kappa\varphi\right)\left[(12n-1)\tan^{2}\left(\kappa\varphi\right)+2\right]}{2\kappa^{2}(n+1)\left|q_{m}^{2}\sec^{2}\left(\kappa\varphi\right)\right|^{3/2}} \,.
\end{equation}
This allows us to explicitly write the function \eqref{W} in the form of the scalar field, similar to the method we used to obtain $V(\varphi)$. In this case, we get that
\begin{align}
    W (\varphi)=&\frac{M2^{n}q_{m}^{2-2n}\left(13\tan^{2}\left(\kappa\varphi\right)-2\right)}{\kappa^{2}}
    \nonumber
    \\
    &
    \times    \left(q_{m}^{2}\sec^{2}\left(\kappa\varphi\right)\right)^{2n-\frac{3}{2}}
    \,.\label{W2Bard}
\end{align}

Now, Eqs. \eqref{L_Bard}  and  \eqref{W2Bard} allow us to write the explicit interaction term for this model:
\begin{align}
W(\varphi)\,\mathcal{L}(F)=&\frac{M2^{n}F^{n+1}q_{m}^{2n-1}\left(13\tan^{2}\left(\kappa\varphi\right)-2\right)}{\kappa^{2}(n+1)}
\nonumber
\\
&
\times\left(\sec\left(\kappa\varphi\right)\right)^{4n-3}\,.
\label{LI_Bard}
\end{align}

Considering $n=0$, Eq. \eqref{LI_Bard}, reduces to
\begin{equation}
W(\varphi)\,\mathcal{L}(F)=    \frac{Mq_{m}^{2}\left(13\tan^{2}(\kappa\varphi)-2\right)F}{\kappa^{2}\left(q_{m}^{2}\sec^{2}(\kappa\varphi)\right)^{3/2}}\, .\label{LI0_Bard}
\end{equation}

\begin{figure}[htb!]
\centering
\includegraphics[scale=0.35]{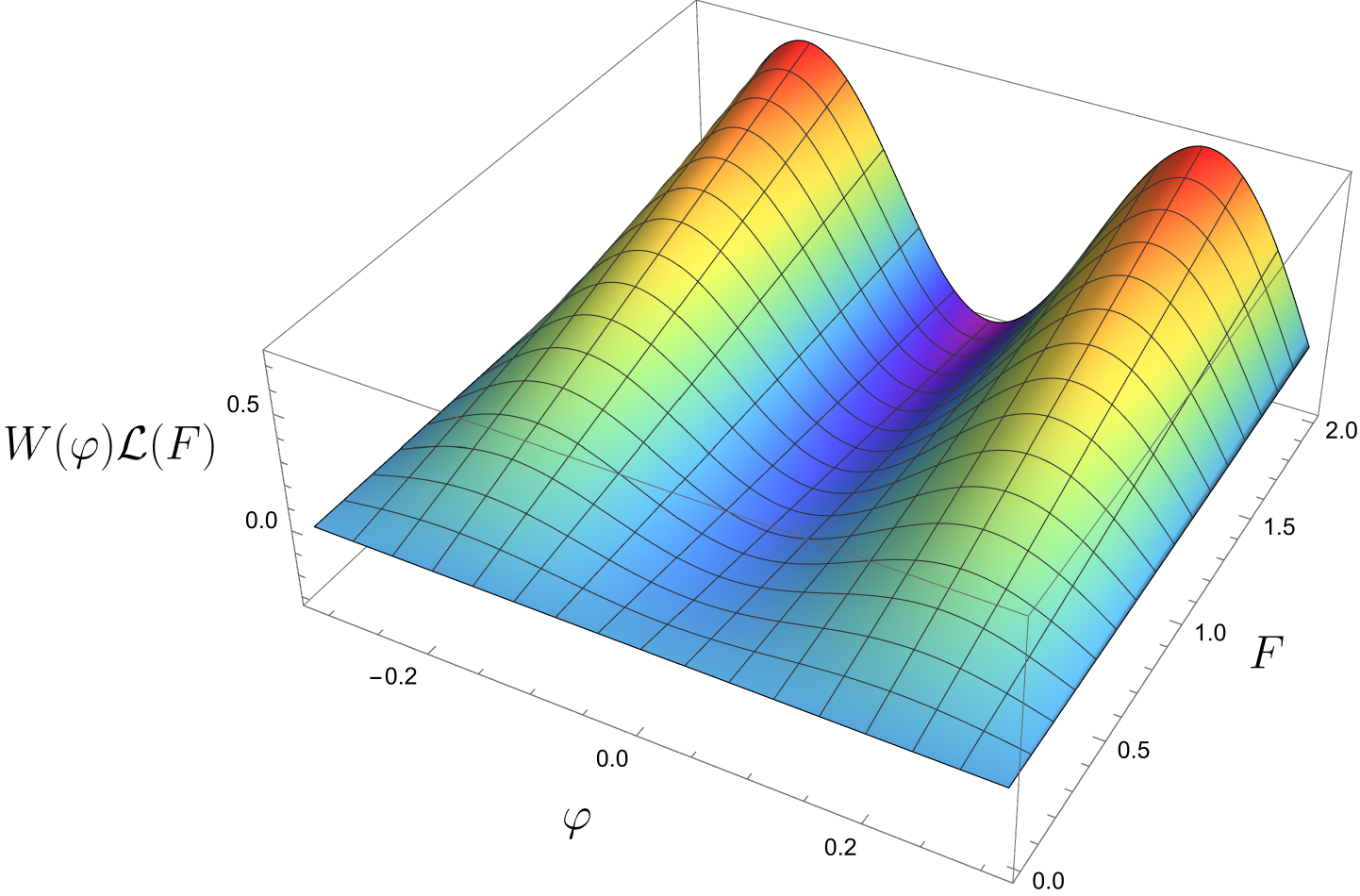} 
\caption{Behavior of the interaction term $W(\varphi)\,\mathcal{L}(F)$ for $n=0$, as described by Eq.~\eqref{LI0_Bard}, with $q_m=0.5$ and $M=1$. The plotting intervals are based on the asymptotic limits of the solution $\varphi(r)$, resulting in $-\frac{\pi}{2\kappa} < \varphi < \frac{\pi}{2\kappa}$, and on the maximum value of $F(r)$ at $r=0$, with $0 \leq F \leq \frac{1}{2q_m^2}$. }
\label{figLI_Bard}
\end{figure}

We illustrate in Fig.~\ref{figLI_Bard} the 3D graph of Eq. \eqref{LI0_Bard}, which reveals that this interaction term $W(\varphi)\mathcal{L}(F)$ has negative values in the region near $\varphi = 0$ (lilac surface) and becomes positive as $|\varphi|$ increases, exhibiting symmetry around this point. For fixed values of $\varphi$, the dependence on $F$ is linear, while two regions of positive maximum appear for higher values of $|\varphi|$. This behavior highlights the transition between negative and positive coupling regimes as the scalar field varies, which indicates the non-trivial nature of the interaction term.
As previously observed (see Fig.~\ref{figBardeenMag}), this transition can be interpreted as the change between a canonical scalar field regime, when $W(\varphi)\mathcal{L}(F) > 0$, and a phantom regime, when $W(\varphi)\mathcal{L}(F) < 0$. This indicates that this term is partially phantom.

%%%%%%%%%%%%%%%%%%%%%%%%%%%%%%%%%%%%%%%%%%%%%%%%%%%%%%%%%%

%%%%%%%%%%%%%%%%%%%%%%%%%%%%%%%%%%%%%%%%%%%%%%%%%%%%

%%%%%%%%%%%%%%%%%%%%%%%%%%%%%%%%%%%%%%%%%%%%%%%%%
\subsubsection{Bardeen model with electric charge: $q_m=0$ and $q_e\neq0$}
%%%%%%%%%%%%%%%%%%%%%%%%%%%%%%%%%%%%%%%%%%%%%%%%%

We now consider the scenario in which only the electric charge is present ($q_e \neq 0$ and $q_m = 0$). In this case, the area functions and the scalar field are given by Eqs. \eqref{sig_qe} and \eqref{phi_qe}, respectively. The metric function \eqref{AMod2} then takes the form
\begin{align}
 A(r)=1-\frac{2Mr^{2}}{\left(q_e^{2}+r^{2}\right)^{3/2}}.
 \label{AMod2qe}
\end{align}

The electromagnetic scalar given by Eq. \eqref{Fe}, associated with this model, is:
\begin{align}
    F=-\frac{q_{e}^{2}\left(2Mq_{e}^{2}-13Mr^{2}\right)^{2}}{2\kappa^{4}\left(q_{e}^{2}+r^{2}\right)^{5}}.\label{Fe_Bard}
\end{align}
This invariant is strictly negative throughout the radial domain, i.e., $F(r)<0$ for all $r\in(-\infty,\infty)$.

To determine the Lagrangian \eqref{Le_BB} explicitly for this model in terms of $r$, we need to determine $W(r)$, since
\begin{align}
  &  {\cal L}(r)=\frac{Mq_{e}^{2}}{\kappa^{2}}\Bigg\{\int\Bigg[\frac{\left(q_{e}^{2}+r^{2}\right)\left(2q_{e}^{2}-7r^{2}\right)W'(r)}{\left(q_{e}^{2}+r^{2}\right)^{9/2}W(r)^{2}}
   \nonumber \\&
    +\frac{4r\left(r^{2}-2q_{e}^{2}\right)}{\left(q_{e}^{2}+r^{2}\right)^{9/2}W(r)}\Bigg]\,dr+\frac{2q_{e}^{2}-7r^{2}}{\left(q_{e}^{2}+r^{2}\right)^{7/2}W(r)}\Bigg\}\,.
\end{align}

Here, similar to the procedure in the previous scenarios, we again model the derivative of the Lagrangian ${\cal L}_{F}(r)$ from a power law, Eq. \eqref{Pot}, but now applied to the expression described by Eq. \eqref{LFe_BB}.
With this, we obtain 
\begin{equation}
   -\frac{\kappa^{2}\left(q_{e}^{2}+r^{2}\right)^{3/2}}{M\left(2q_{e}^{2}-13r^{2}\right)W(r)}=\alpha F^n \,,
\end{equation}
whcih yields the following solution for $W(r)$
\begin{equation}
   W(r)=-\frac{(-2)^{n}\kappa^{4n+2}q_{e}^{-2n}\left(q_{e}^{2}+r^{2}\right)^{5n+\frac{3}{2}}}{\alpha\left[M\left(2q_{e}^{2}-13r^{2}\right)\right]^{2n+1}}\,.
   \label{We_Bard}
\end{equation}

From the solution \eqref{We_Bard}, we find that the electromagnetic quantities, the term $W(r){\cal L}(r)$ and  potential $V(r)$, are now described as
\begin{align}
    {\cal L}(r)&=-\frac{(-1)^n\alpha q_{e}^{2n+2}\left[M\left(2q_{e}^{2}-13r^{2}\right)\right]^{2(n+1)}}{2^{n+1}(n+1)\kappa^{4(n+1)}\left(q_{e}^{2}+r^{2}\right)^{5(n+1)}},%\label{Le_Bard}
    \nonumber\\
    {\cal L}_F(r)&=\frac{\alpha(-2)^{-n}q_{e}^{2n}\left[M\left(2q_{e}^{2}-13r^{2}\right)\right]^{2n}}{\kappa^{4n}\left(q_{e}^{2}+r^{2}\right)^{5n}}\,,
   \nonumber \\
   W(r){\cal L}(r) &=\frac{Mq_{e}^{2}\left(2q_{e}^{2}-13r^{2}\right)}{2\kappa^{2}(n+1)\left(q_{e}^{2}+r^{2}\right)^{7/2}}\,,
   \nonumber\\
    V (r)&=-\frac{M (2 n-1) q_e^2}{2 \kappa ^2 (n+1) \left(q_e^2+r^2\right)^{5/2}} \,,\label{Ve2_Bar}
\end{align}
respectively.

Considering $n=0$ and $\alpha=1$, we obtain in this particular case, the following solutions
\begin{align}    W(r)&=\frac{\kappa^{2}\left(q_{e}^{2}+r^{2}\right)^{3/2}}{13Mr^{2}-2Mq_{e}^{2}},%\label{We0_Bard}
     \nonumber\\
    {\cal L}(r)&=-\frac{M^{2}q_{e}^{2}\left(2q_{e}^{2}-13r^{2}\right)^{2}}{2\kappa^{4}\left(q_{e}^{2}+r^{2}\right)^{5}},%\label{Le0_Bard}
    \nonumber\\
    {\cal L}_F(r)&=1 \,,
    \nonumber\\
   W(r){\cal L}(r) &=\frac{Mq_{e}^{2}\left(2q_{e}^{2}-13r^{2}\right)}{2\kappa^{2}\left(q_{e}^{2}+r^{2}\right)^{7/2}} \,,
   \nonumber\\
    V (r)&=\frac{Mq_{e}^{2}\left(2q_{e}^{2}-r^{2}\right)}{2\kappa^{2}\left(q_{e}^{2}+r^{2}\right)^{7/2}} \,.\label{Ve2n0_Bardeen}
\end{align}
Note that for this value of $n$, the Lagrangian in \eqref{Ve2n0_Bardeen} is identical to the electromagnetic scalar \eqref{Fe_Bard}.  Therefore, in this specific case where $n=0$, we have a linear Lagrangian ${\cal L}(F)\equiv F$.

As shown throughout the manuscript, the relevant quantities, namely, ${\cal L}(r)$, $W(r){\cal L}(r)$, and $V(r)$, depend on the function $W(r)$, defined in this model by Eq.~\eqref{We_Bard}, in order to be expressed analytically. However, this specific function exhibits a limitation when the radial coordinate approaches $r \to \pm \sqrt{2/13}\, q_e$. Nevertheless, when considering this limit in the expressions of the physical quantities that can be assigned to observations, as presented in Eqs.\eqref{Ve2_Bar} and \eqref{Ve2n0_Bardeen} for the case $n = 0$, we do not identify any inconsistency.

We also verify that the quantities $\rho(r)$, $p_{r}(r)$ and $p_{t}(r)$, which result from the relations $\rho(r) = T^{t}_{\phantom{t} t}$, $p_r(r) = -T^{r}_{\phantom {r} r}$ and $p_t(r) = -T^{\theta}_{\phantom{\theta} \theta}$, or, equivalently, as in Sec.~\ref{EC}, as well as the electric field $E(r)$, also remain divergence-free at this point. The corresponding expressions are given by:
\begin{align}
    \rho (r)  &=\frac{q_{e}^{2}}{\kappa^{2}}\left(\frac{8Mr^{2}}{\left(q_{e}^{2}+r^{2}\right)^{7/2}}-\frac{1}{\left(q_{e}^{2}+r^{2}\right)^{2}}\right),
     % \label{eq:rhoe}
    \nonumber \\
 p_r(r)   &= -\frac{q_{e}^{2}\left[4Mr^{2}+\left(q_{e}^{2}+r^{2}\right)^{3/2}\right]}{\kappa^{2}\left(q_{e}^{2}+r^{2}\right)^{7/2}},
     % \label{eq:pre}
    \nonumber \\
    p_t(r)  &= \frac{q_{e}^{2}\left[-2Mq_{e}^{2}+5Mr^{2}+\left(q_{e}^{2}+r^{2}\right)^{3/2}\right]}{\kappa^{2}\left(q_{e}^{2}+r^{2}\right)^{7/2}},
    \label{eq:pte}
   \nonumber \\
    E(r) & = \frac{Mq_{e}\left(2q_{e}^{2}-13r^{2}\right)}{\kappa^{2}\left(q_{e}^{2}+r^{2}\right)^{5/2}}. 
\end{align}
These expressions, show that all physical quantities, as represented in Eqs.\;\eqref{Ve2_Bar}--\eqref{eq:pte}, remain finite both at the limit $r \to 0$ and at $r \to \infty$, as well as at the specific value $r \to \pm \sqrt{2/13} \,q_e$. Despite the limitation associated with the function $W(r)$, this limitation does not affect the consistency or validity of the proposed model.

In this model, it was not possible to obtain $r(F)$, but we were able to find $r(\varphi)$. Therefore, we express only $V(\varphi)$ and $W(\varphi)$, respectively, as
\begin{align}
       V (\varphi)&= -\frac{M\cos^{4}(\kappa\varphi)\left[(14n+1)\tan^{2}(\kappa\varphi)-4n-2\right]}{2\kappa^{2}(n+1)\left(q_{e}^{2}\sec^{2}(\kappa\varphi)\right)^{3/2}}\, ,\label{Ve2_Bard}
       \\
       W(\varphi)&=-\frac{(-2)^{n}\kappa^{4n+2}q_{e}^{-2n}\left(q_{e}^{2}\sec^{2}(\kappa\varphi)\right)^{5n+\frac{3}{2}}}{\alpha\left[Mq_{e}^{2}\left(2-13\tan^{2}(\kappa\varphi)\right)\right]^{2n+1}}.\label{We2_Bard}
\end{align}

Whereas these quantities, $V(\varphi)$ and $W(\varphi)$, respectively, with $n=0$ are given by:
\begin{align}
       V (\varphi)&= -\frac{M\cos^{4}(\kappa\varphi)\left(\tan^{2}(\kappa\varphi)-2\right)}{2\kappa^{2}\left(q_{e}^{2}\sec^{2}(\kappa\varphi)\right)^{3/2}} \,,
       \label{Ve3_Bard}
       \\
       W(\varphi)&= \frac{\kappa^{2}\left(q_{e}^{2}\sec^{2}(\kappa\varphi)\right)^{3/2}}{Mq_{e}^{2}\left(13\tan^{2}(\kappa\varphi)-2\right)}.\label{We3_Bard}
\end{align}

%%%%%%%%%%%%%%%%%%%%%%%%%%%%%%%%%%%%%%%%%%%%%%%%%%%%%%%%%%

\subsubsection{Kretschmann scalar}

The Kretschmann scalar associated with the metric \eqref{AMod2} is obtained from Eq.~\eqref{K} and can be written explicitly as
\begin{align}
    K=&\frac{4}{\left(a^{2}+r^{2}\right)^{7}}\bigg[8Ma^{2}r^{2}\left(-2a^{4}-a^{2}r^{2}+r^{4}\right)\sqrt{a^{2}+r^{2}}
    \nonumber\\
    &    +M^{2}\left(4a^{8}-44a^{6}r^{2}+169a^{4}r^{4}-68a^{2}r^{6}+12r^{8}\right)
    \nonumber\\
    &    +3a^{4}\left(a^{2}+r^{2}\right)^{3}\bigg]\, .
    \, \label{K2}
\end{align}

At the central limit, $r\to0$, the scalar assumes a finite value that depends only on the parameters of the solution:
\begin{equation}
\lim_{r\to0}K=\frac{4 \left(4 M^2+3 a^2\right)}{a^6}\,.
\end{equation}
On the other hand, for very large distances ($r\to\infty$), we have
\begin{equation}
\lim_{r\to\infty}K(r)=0,
\end{equation}
which confirms the regularity of the spacetime and its convergence to the asymptotically flat regime, as expected for BB solutions.

%%%%%%%%%%%%%%%%%%%%%%%%%%%%%%%%%%%%%%%%%%%%%%%%%
\section{Energy conditions}\label{EC}
%%%%%%%%%%%%%%%%%%%%%%%%%%%%%%%%%%%%%%%%%%%%%%%%%%%%%%%%%%

To conclude this article, let us briefly discuss how the energy conditions can be analyzed when considering both the magnetic and electric cases. To do this, we identify the components of the energy-momentum tensor as described by a fluid. In the region outside the horizon, where $A(r)>0$, the following holds
\begin{equation}
T_{\phantom{\mu}\nu}^{\mu}=W(\varphi)\,\overset{F}{T}{}_{\phantom{\mu}\nu}^{\mu}+\overset{\varphi}{T}{}_{\phantom{\mu}\nu}^{\mu}=\text{diag}\left(\rho,\,-p_{r},\,-p_{t},\,-p_{t}\right),\label{Comp_TA}
\end{equation}
while for $A(r)>0$, they are given as
\begin{equation}
T_{\phantom{\mu}\nu}^{\mu}=W(\varphi)\,\overset{F}{T}{}_{\phantom{\mu}\nu}^{\mu}+\overset{\varphi}{T}{}_{\phantom{\mu}\nu}^{\mu}=\text{diag}\left(-p_{r},\,\rho,\,-p_{t},\,-p_{t}\right).\label{Comp_TA2}
\end{equation}
With this, the following inequalities express the energy conditions for the energy-momentum tensor \eqref{Comp_TA} as:
\begin{align}
{\rm NEC}_{r,t}^{q_m,q_e}&=\rho+p_{r,t}\geq 0\,,\label{NEC}\\
{\rm SEC}_{(rt)}^{q_m,q_e}&=\rho + p_{r} + 2p_t\geq 0\,,\label{SEC}\\
{\rm DEC}_{r,t}^{q_m,q_e}&=\rho \, - \mid p_{r,t}\mid \geq 0\, \quad\text{or}\quad \rho\pm p_{r,t}\geq 0\,,\label{DEC}\\
{\rm DEC}_{}^{q_m,q_e}&=\rho\geq 0\,,\label{WEC}
\end{align}
where NEC, SEC, DEC, denote the null, strong and dominant energy conditions.
In these expressions, the indices $r$ and $t$ refer, to the radial and tangential components of the anisotropic fluid respectively. The indices $q_m$ and $q_e$ indicate whether it is a case in which only the magnetic charge or only the electric charge is involved.

By explicitly evaluating \eqref{NEC}–\eqref{WEC} for the Simpson–Visser \eqref{AMod1qm}–\eqref{AMod1qe} and Bardeen \eqref{AMod2qm}–\eqref{AMod2qe} metrics, we verify that the obtained forms are in agreement with those already presented in previous studies (see, e.g. \cite{Rodrigues2023,Alencar:2024yvh}. It is worth noting that this result was obtained without specifying any value for the constant $n$.

%%%%%%%%%%%%%%%%%%%%%%%%%%%%%%%%%%%%%%%%%%%%%%%%%%%%%%%%%%%%%%%%
\section{Summary and Conclusion}\label{sec:concl}
%%%%%%%%%%%%%%%%%%%%%%%%%%%%%%%%%%%%%%%%%%%%%%%%%%%%%%%%%%%%%%%%

In this work, we investigated BB solutions within the framework of GR, restricting our analysis to static and spherically symmetric geometries. The central objective of our study was to explore how such spacetimes can be consistently supported by a combination of a scalar field and NLED, while further incorporating a nonminimal coupling term between these two matter sectors, thus extending the usual minimal coupling approach.  
To this end, in addition to the standard Lagrangian associated solely with the scalar field, we introduced an explicit interaction term of the form $W(\varphi)\mathcal{L}(F)$, where the scalar field $\varphi$ effectively modulates the contribution of the electromagnetic sector. This framework allowed us to investigate, in a systematic manner, both the purely magnetic case ($q_m \neq 0$, $q_e = 0$) and the purely electric case ($q_m = 0$, $q_e \neq 0$). For each scenario, we solved the field equations and reconstructed the relevant matter functions, namely ${\cal L}(r)$, ${\cal L}_F(r)$, $W(r)$, $\epsilon(r)$, and the scalar potential $V(r)$, which together ensure the consistency and regularity of the solutions.

Our analysis encompassed two well-known classes of regular spacetimes, namely the Simpson--Visser and Bardeen BB models. For each case, we explicitly reconstructed the corresponding matter functions, demonstrating that, by assuming a power-law behavior for ${\cal L}_F(r)$, it becomes possible to derive closed analytical expressions for $W(r)$ and, consequently, for ${\cal L}(r)$. This procedure established a systematic framework to characterize radial profiles of coupling functions, scalar potentials, and electromagnetic invariants, thereby deepening our understanding of the matter sector sustaining these geometries.
The results show that the reconstruction method consistently reproduces known solutions (for specific limits of $n$ and $\alpha$) while naturally extending them to a broader class of regular configurations. A particularly notable outcome of this approach is that it allows for the construction of solutions supported by linear electrodynamics (LED), in contrast to the more conventional frameworks relying on NLED. This constitutes one of the central results of the present work: by appropriately fixing the model parameters (specifically, by setting $n=0$), we have shown that it is possible to obtain BB solutions without invoking NLED, thereby challenging the widespread expectation that BH regularization necessarily requires nonlinear matter sources.

Furthermore, for completeness, we verified that the matter sector derived from our action~\eqref{action} coincides with the standard matter action usually associated with BB solutions in the literature, both in the purely magnetic and purely electric cases. In other words, when considering only the matter Lagrangians of the Simpson--Visser and Bardeen models, we recover the same contribution by appropriately assigning $q_m$ or $q_e$, as summarized below:
\begin{equation}
	\mathcal{L}_{\rm \varphi(usual)}(\varphi)+\mathcal{L}_{\rm (usual)}(F)
	=\mathcal{L}_{\varphi}(\varphi)+W(\varphi)\mathcal{L}(F)\,.
\end{equation}
We therefore conclude that the total matter Lagrangian, obtained as the sum of the scalar field contribution, Eq.~\eqref{Lphi}, and the electromagnetic contribution weighted by the interaction factor $W(\varphi)$, Eq.~\eqref{LI}, exactly reproduces the standard Lagrangian of black-bounce solutions supported by NLED~\cite{Rodrigues2023,Alencar:2024yvh}, for both magnetic and electric configurations. Thus, this equivalence emerges naturally from the formalism developed throughout this work, without requiring the imposition of specific values for the parameter $n$. Hence, our analysis suggests that this identity is valid in a general setting and should consistently hold across arbitrary BB models.

In addition to satisfying the energy conditions, theories of nonlinear electrodynamics (NLED) are subject to several important conceptual limitations. In particular, they may violate fundamental criteria such as causality and unitarity in certain regions of spacetime, especially near the regular center. Another significant issue arises in electrically charged configurations, where the corresponding Lagrangian can become multivalued, thereby compromising its physical consistency. Moreover, NLED-based solutions are sometimes prone to instabilities under various types of perturbations.  
These challenges highlight the need to explore alternative approaches capable of mitigating such problems. In this context, a description based on LED provides a more robust and physically coherent framework, offering a conceptually simpler and more tractable setting for investigating regular solutions in extreme gravitational regimes.

In summary, this work presents a systematic framework that deepens our understanding of regular BB solutions supported by scalar fields and NLED. By reconstructing the matter functions and exploring both magnetic and electric configurations, we have demonstrated the versatility of this approach and highlighted the potential of LED as an alternative to conventional NLED constructions.  
This framework opens several avenues for future investigations. Notably, it motivates: (i) detailed perturbative stability analyses to assess the robustness of these geometries under various types of fluctuations; (ii) extensions to dynamical or non-static scenarios, including rotating or time-dependent solutions; and (iii) the incorporation of additional contributions to the action, such as cosmological terms, higher-curvature corrections, or generalized scalar couplings. Beyond these directions, this approach may inspire studies of astrophysical observables—BH shadows, gravitational waves, and matter near regularized compact objects—thus bridging theory and phenomenology.

%\vspace{1cm}

%%%%%%%%%%%%%%%%%%%%%%%%%%%%%%%%%%%%%%%%%%%%%%%%%%%%%%%%%%%%%%%%
\acknowledgments{MER thanks Conselho Nacional de Desenvolvimento Cient\'ifico e Tecnol\'ogico - CNPq, Brazil, for partial financial support. This study was financed in part by the Coordena\c{c}\~{a}o de Aperfei\c{c}oamento de Pessoal de N\'{i}vel Superior - Brasil (CAPES) - Finance Code 001.
FSNL acknowledges support from the Funda\c{c}\~{a}o para a Ci\^{e}ncia e a Tecnologia (FCT) Scientific Employment Stimulus contract with reference CEECINST/00032/2018, and funding through the research grants UIDB/04434/2020, UIDP/04434/2020 and PTDC/FIS-AST/0054/2021.}
%%%%%%%%%%%%%%%%%%%%%%%%%%%%%%%%%%%%%%%%%%%%%%%%%%%%%%%%%%%%%%%%

%%%%%%%%%%%%%%%%%%%%%%%%%%%%%%%%%%%%%%%%%%%%%%%%%%%%%%%%%%%%%%%%

%%%%%%%%%%%%%%%%%%%%%%%%%%%%%%%%%%%%%%%%%%%%%%%%%%%%%%%%%%%%%%%%

%%%%%%%%%%%%%%%%%%%%%%%%%%%%%%%%%%%%%%%%%%%%%%%%%%%%%%%%%%%%%%%%
\end{document}